\documentclass[conference]{IEEEtran}
\IEEEoverridecommandlockouts
\usepackage{cite}
\usepackage{amsmath,amssymb,amsfonts}
\usepackage{algorithmic}
\usepackage{textcomp}
\usepackage{xcolor}
\usepackage[pdftex]{graphicx}
\usepackage{epstopdf}

\usepackage{amsmath,amssymb,enumerate} 
\usepackage{bm} 
\usepackage{indentfirst} 
\usepackage{empheq} 
\usepackage{bm}
\usepackage{comment}
\usepackage{here} 
\newcommand{\dd}{\mathrm{d}}
\newcommand{\e}{\mathrm{e}}
\newcommand{\ii}{\mathrm{i}}

\def\BibTeX{{\rm B\kern-.05em{\sc i\kern-.025em b}\kern-.08em
    T\kern-.1667em\lower.7ex\hbox{E}\kern-.125emX}}
\begin{document}

\title{Evaluation of User Dynamics Created by Weak Ties among Divided Communities
}

\author{\IEEEauthorblockN{Takahiro Kubo}
    \IEEEauthorblockA{
        \textit{Tokyo Metropolitan University} \\
Tokyo 191--0065, Japan \\
kubo-takahiro@ed.tmu.ac.jp}
\and
\IEEEauthorblockN{Chisa Takano}
    \IEEEauthorblockA{
        \textit{Hiroshima City University} \\
Hiroshima 731-3194, Japan \\
takano@hiroshima-cu.ac.jp}
\and
\IEEEauthorblockN{Masaki Aida}
    \IEEEauthorblockA{
        \textit{Tokyo Metropolitan University} \\
Tokyo 191--0065, Japan \\
aida@tmu.ac.jp}
}

\maketitle

\begin{abstract}
Flaming phenomena represent the divergence in the strength of user dynamics as created by user interactions in online social networks (OSNs). 
Although it has been known that flaming phenomena occur when the Laplacian matrix of the OSN has non-real eigenvalues, it was recently shown that flaming phenomena may occur even if all the eigenvalues are real numbers.
This effect appears only in the situation that some eigenvalues are degenerate, and a special unitary transformation is applied to the equations representing user dynamics; whether actual OSNs satisfy this condition has not been fully discussed.
In this paper, we clarify that the user dynamics caused by the degeneration of eigenvalue $0$ is one specific example of the above condition. 
We also investigate the mechanism and characteristics of flaming phenomena generated by degenerated eigenvalues.
Furthermore, we demonstrate through numerical simulations that the degeneration of eigenvalues can cause divergence.
\end{abstract}

\begin{IEEEkeywords}
online social network, wave equation, flaming, phase synchronization, Kuramoto model
\end{IEEEkeywords}

\section{Introduction}
The development and spread of information networks have dramatically activated interpersonal communication and the ability of individuals to disseminate information; they have revolutionized the way we do business and live our daily lives.
In particular, various social networking services (SNSs) provided on the Internet support the activation of social networks.
We call the relationship between such users an online social network (OSN).
Activation of OSNs is the foundation for highly efficient social activities, and for this reason, SNSs are indispensable in modern society.
However, we cannot ignore the fact that OSNs also have the risk of causing social instability.
The existence of SNSs has played a highly negative role in international terrorist activities and the intensification of social divisions based on ideological, political, and religious disagreements.
While information network technologies contribute to the efficiency of social activities, they can also accelerate the destabilization of social systems.

Typical of the negative impacts on real-world activities possible is the flaming phenomena.
In~\cite{Alonzo_2004},\cite{Peter_2010}, the definition of flaming phenomena is drawn from observations of actual network services. 
For example, \cite{Peter_2010} defines flaming phenomena as `displaying hostility by insulting, swearing or using otherwise offensive language' from observations of comments generated in an actual YouTube service.
In such situations, it leads to the activation of user discussions and/or the escalation of user conflicts.
Our definition of flaming phenomena, based on an analytical model of user dynamics, does not contradict the observation-based definition.
The activation or escalation can be represented as phenomena that increases the intensity of user dynamics.
Therefore, in our framework, the definition of the flaming phenomena is that the strength of user dynamics on OSNs diverges as users interact with each other.
It is difficult to provide an immediate and fundamental solution to the flaming phenomena because their causes are rooted in various social characteristics and problems.
However, if we leave the phenomena until a full solution is found, there is a risk that the spread of flaming phenomena will yield severely negative social effects.
Therefore, it is desirable to develop mitigation techniques for flaming phenomena that can be implemented immediately, without addressing the underlying social causes.
It is desirable to understand the system's characteristics common to flaming phenomena and to consider countermeasures, rather than trying to analyze the specific factors that cause each case of flaming.
A model proposed to understand the engineering mechanism of flaming phenomena that involves OSN structure is the oscillation model~\cite{aida_2018}.
This model describes user dynamics in terms of the wave equations defined on the OSN, and can not only describe user dynamics as a solution to the wave equation but also explicitly describe the causal relationship between the structure of the OSN and user dynamics.

It is known that the network oscillation model predicts that flaming phenomena occur when the Laplacian matrix, which represents the structure of OSN, has non-real eigenvalues.
This enables us to understand the causes of flaming phenomena in relation to the network structure and provides a framework for considering countermeasures to flaming phenomena within the framework of network control technology.
The conventional understanding is that flaming phenomena could not occur when all eigenvalues of the Laplacian matrix representing the OSN structure were real numbers.

Recently, however, \cite{kubo_2019} showed that even if all eigenvalues of the Laplacian matrix representing the OSN structure were real numbers, flaming phenomena can still occur under the following two conditions.
\begin{itemize}
\item When the Laplacian matrix cannot be diagonalized.\\
The prerequisite for the inability to diagonalize the Laplacian matrix is that some of the eigenvalues of the Laplacian matrix must be degenerate.
However, different from a purely mathematical model, it is extremely unlikely that the eigenvalues would be degenerate in a Laplacian matrix of a complex network structure, such as the structure of an OSN.
This corresponds to the fact that the eigenvalue equation has multiple roots.
Considering that, in general, it is extremely unlikely that a $n$-th order equation with randomly determined coefficients will have multiple solutions, we can understand that this is a condition in which duplication of eigenvalues is extremely unlikely.
However, there is an exception wherein duplication of eigenvalues can readily occur, which we discuss later.
\item An equation with a specific unitary transformation is used instead of the original equation to describe user dynamics.\\
In the oscillation model, the user dynamics are described by wave equations defined on the network.
This (original) wave equation is invariant to the unitary transformation.
However, if we extend the wave equation to explicitly describe the causal relationship between the network structure and user dynamics, the extended equations are not invariant to unitary transformations.
This means that we have different extended equations depending on whether the starting point is the original wave equation or the unitary-transformed wave equation. 
The user dynamics derived using the unitary-transformed equation may diverge even when all eigenvalues of the Laplacian matrix are real numbers.
However, the unitary-transformed equation is different from the original wave equation and therefore is generally different from the phenomena that occur in the real world.
However, there is an exception with regard to the extended equations being invariant for the unitary transformation.
We will discuss its user dynamics later.
\end{itemize}

In social network analysis, there are many examples of relationships that are not usually associated with each other, but sometimes provide important ``insights'' that lead to advances in the future.
For example, a person with whom one does not normally have much contact may bring unexpected information about a new job, or an exchange between researchers in different fields may provide a solution to an intractable problem.
Such a link, which is not strong, but does play an important role, is called a ``weak tie''.
From this point of view, the phenomenon that occurs when normally separated communities are linked for some reason is an interesting subject of study with regard to the user dynamics of social networks.
It is known that the Laplacian matrix representing the network structure always has an eigenvalue of $0$, and the degenerate number of eigenvalue $0$ is equal to the number of connected components of the network.
Hence, for an entire connected network, there is only one eigenvalue $0$ in the Laplacian matrix representing its structure, but for a network split into two parts, there are two duplicate eigenvalue $0$ instances in the Laplacian matrix representing its structure.
As mentioned in the first bullet point above, the degeneration of eigenvalues of the Laplacian matrix is a very unlikely situation for eigenvalues other than $0$.
However, it is not uncommon for overlap to occur in terms of eigenvalue $0$; it represents a split into multiple isolated networks.
This corresponds to the emergence of a fragmented community in OSNs.
Furthermore, in relation to the second bullet point above, there is only one case where the equation explicitly describes the causal relationship with the original equation, even with the unitary transformation, and that is when we consider a part of the equation corresponding to zero eigenvalues.
Thus, user dynamics in a situation where some changes occur in divided communities that are linked by ``weak ties'' can be considered to correspond to the typical case for the situation assumed in the literature \cite{kubo_2019}.

In this paper, we consider user dynamics generated from weak interactions between divided communities as one specific example of the mechanism of flaming phenomena shown in~\cite{kubo_2019}.
Specifically, we study the characteristics of the user dynamics generated by the interaction of the oscillation modes, which are degenerate with eigenvalue $0$ and interact with each other in Jordan canonical form.
This paper is organized as follows. 
Section 2 presents related research on the user dynamics of OSNs and clarifies the position and importance of this research.
Section 3 describes the modeling of user dynamics arising from the weak interaction between divided communities and examines the properties of user dynamics that arise from them.
In Section 4, we show the validity of the flaming phenomena derived in Section 3 by means of numerical simulations.
Finally, we conclude this paper in Sec. 5.

\section{Related Work}
Since it is easier to obtain structural data for online social networks than real-world social networks, research on network analysis has been very active~\cite{Watts2007,Borgatti2009,Scott2011}.
Online social networks can be roughly classified into two types: social relationship networks~\cite{Liu2005,Catanese2011,Mislove2008,Ahn2007,Leskovec2010}, which are formed by the relationship between followers, and social interaction networks~\cite{Holger2002,Kossinets2006,Onnela2007,Isella2011}, which are the actual interactions between peers.
To deeply understand the structure of online social networks, it is necessary to apply theoretical models in addition to analyzing real data. 

Studies on theoretical models of online social networks can be summarized from the following points of view. 

There are two main models for explaining the diffusion of a new service or the spread of a rumor on an OSN. 
One is macroscopic modeling, which does not directly describe the state of individual users, while the other is microscopic modeling, which deals directly with the state of each individual user.
The macroscopic model focuses on the size of the set of users with specific characteristics, such as the number of service subscribers or the number of people who know about the rumor, while the microscopic model focuses on the state of individual users~\cite{Song2007,Leenders1995}. 

With regard to the macroscopic model, studies have detailed the SIS and SIR models of infectious diseases, in which service subscribers and people who are aware of rumors are likened to those who are infected with an infectious disease~\cite{PeiMakse2013,Hethcote2000,May2001,Pastor-Satorras2001}.
These models describe the dynamics of infection by representing the time variation of macroscopic quantities in the network as first-order differential equations. 
The SIS and SIR models do not model the dynamics of individual users and do not need to be aware of the microscopic state of individual users.
In these models, explosive user dynamics refers to a high rate of state change (e.g., rapid propagation of rumors) and does not deal with the divergence of the strength of user dynamics, e.g. flaming phenomena.

As microscopic models, studies using percolation models have described individual user dynamics~\cite{Sorin2000,Martin2008,Campbell2013,Morone2015}.
These models show how the state change spills over to the whole system through the dynamics of individual user state transitions.
As well as percolation, the Markov chain model is known as a model that can deal with the dynamics of individual user state~\cite{Song2007,Leenders1995,Watts2007b,Galuba2010,Barbieri2013,Mieghem2014}.
Song et al. modeled the diffusion of innovation in the network (NW) using continuous-time Markov chains to adjust the speed of flow between individuals to account for communication delay~\cite{Song2007}.
On the other hand, Leenders et al. expanded the possibilities for longitudinal binary social network data to be treated as continuous-time Markov chains~\cite{Leenders1995}.
Modeling based on continuous-time Markov chains can describe individual user state changes, but it cannot describe the divergence of user dynamics, such as flaming phenomena, because it considers only the steady-state condition and the transient state leading to it.

In this study, we deal with wave equations on networks~\cite{aida_2018,Aida2016,Network_Dynamics_Introduction}.
This reflects the fact that the influence between OSN users should propagate at a finite velocity.
The model is capable of representing the state of individual users and describing the divergence in the strength of user dynamics, such as flaming phenomena.

\section{Degenerated Oscillation Model in Multiple Isolated Networks}
\subsection{User Dynamics When the Eigenvalues Degenerate}
In this subsection, we introduce the oscillation model on the network that describes how users interact with each other on an OSN.

The oscillation model on networks is a minimal model for describing the user's state. 
It also rules of interaction between users, as simply and as universal as possible.
We describe the state of a user by a one-dimensional variable, $x_i(t)$, which is the state of user $i$ at time $t$, where $i=1, \,2, \, \dots, \,n$ and $n$ is the number of users.
The definition of the state vector for all users is written as 
\begin{align}
\bm{x}(t) := {}^t\!(x_1(t),\,\dots,\,x_n(t)), 
\end{align}
where superscript $t$ represents the transposition of the state vector. 
Let us consider the interaction between users.
User $i$ receives a force, $F_{ij}$, from neighboring user $j$ that makes $|x_i(t) - x_j(t)|$ smaller.
The strength of the force is proportional to $x_i(t) - x_j(t)$ and is written as follows
\begin{align}
F_{ij} = -w_{ij} \, (x_i(t) - x_j(t)),
\label{eq:RestoringF}
\end{align}
where $w_{ij} \ge 0$ is a constant.
Since the interaction between users is generally asymmetric, $w_{ij} \not= w_{ji}$.
Then, the equation of motion of the state vector of users is written as the following wave equation: 
\begin{align}
\frac{\dd^2}{\dd t^2}\,\bm{x}(t) = -\bm{\mathcal{L}}\,\bm{x}(t),
\label{eq:EoM}
\end{align}
where $\bm{\mathcal{L}}$ is the Laplacian matrix of the directed graph whose weight of directed link 
from $i$ to $j$ is given by $w_{ij}$.
Although several definitions of the Laplacian matrix are known, we follow the definition in~\cite{aida_2018}.
Note that we assume all the eigenvalues of $\bm{\mathcal{L}}$ are realnumbers because we concentrate the discussion on the generation of flaming phenomena under this condition. 

First, let us consider the case where Laplacian matrix $\bm{\mathcal{L}}$ is diagonalizable.
By applying a transformation such that $\bm{\mathcal{L}}$ is diagonalized,
we transform the wave equation (\ref{eq:EoM}) into 
\begin{align}
\frac{\dd^2}{\dd t^2}\,\bm{\phi}(t) = -\bm{\Lambda}\,\bm{\phi}(t),
\label{eq:EoM2}
\end{align}
where $\bm{\Lambda}$ is the diagonalized Laplacian matrix, which is $\bm{\Lambda} := \bm{P}^{-1}\,\bm{\mathcal{L}}\,\bm{P}$ given the appropriate square matrix $\bm{P}$.
In addition, $\bm{\phi}(t) := \bm{P}^{-1}\, \bm{x}(t)$.
The wave equation (\ref{eq:EoM}) on OSNs can be decomposed into $n$ independent equations for oscillation modes. 
Then, the solutions of (\ref{eq:EoM}) are given as follows:
\begin{align}
\bm{\phi}(t)=\exp[\mp\ii\, \bm{\Omega}\, t]\, \bm{\phi}(0),
\label{eq:Solution0}
\end{align}
where, for the eigenvalues of Laplacian matrix $\lambda_\mu$ $(\mu=0,\,1,\,\dots,\,n-1)$, $\Omega$ can be defined as
\[
\bm{\Omega} := \mathrm{diag}\!\left(\sqrt{\lambda_0},\,\sqrt{\lambda_1},\,\dots,\,\sqrt{\lambda_{n-1}}\right).
\]

If $\bm{\mathcal{L}}$ is not diagonalizable, then the wave equation (\ref{eq:EoM}) is transformed into the form of (\ref{eq:EoM2}) by using the Jordan canonical form of $\bm{\mathcal{L}}$, $\bm{\Lambda}$.
The transformed wave equation (\ref{eq:EoM2}) using Jordan canonical form cannot be decomposed into independent oscillation modes due to the effect of the non-diagonal component of $\bm{\Lambda}$.
Therefore, in order to clarify the influence of the non-diagonal component, we introduce the fundamental equation of networks~\cite{aida_2018,Network_Dynamics_Introduction}.

First, let us introduce an $n\times n$ matrix $\bm{H}$ that satisfies
\begin{align}
\bm{H}^2=\bm{\mathcal{L}}.
\label{eq:H2Lambda}
\end{align}
Then, we can write the fundamental equation of (\ref{eq:EoM}) as
\begin{align}
\pm\ii\,\frac{\dd \, \bm{x}(t)}{\dd t} = \bm{H}\,\bm{x}(t).
\label{eq:EoM3}
\end{align}
The solution of (\ref{eq:EoM3}) also satisfies the original wave equation (\ref{eq:EoM}).
If $\bm{\mathcal{L}}$ is not diagonalizable, then $\bm{H}$ is not diagonalizable either.
For this reason, we transform the fundamental equation (\ref{eq:EoM3}) so that $\bm{H}$ is represented in Jordan canonical form $\bm{\Omega}$ as follows
\begin{align}
\pm\ii\,\frac{\dd \, \bm{\phi}(t)}{\dd t} = \bm{\Omega}\,\bm{\phi}(t).
\label{eq:EoM4}
\end{align}
In (\ref{eq:EoM4}), $\bm{\Omega}$ is expressed as follows
\begin{align}
	\bm{\bm{\Omega}} = \begin{bmatrix}
 \ddots & & & & &\\
 & \omega_\mu & 1 & & \mbox{\Huge 0} &\\
 &  & \omega_\mu & 1 & &\\
 &  &  & \ddots & 1 &\\
 & $\mbox{\Huge 0}$ &  &  & \omega_\mu &  &\\
 &  &  &  &  & \ddots & \\
\end{bmatrix}, 
\label{eq:JordanCanonicalForm}
\end{align}
where diagonal component $\omega_{\mu}$ represents an oscillation mode.
Let us focus on the example of the Jordan block of oscillation mode $\omega_{\mu}$, shown below, as it cannot be divided further.
\begin{align}
	\bm{\Omega}_{\mathrm{J}} = \begin{bmatrix}
\omega_{\mu} & 1 & 0 \\
0 & \omega_{\mu} & 1\\
0 & 0 & \omega_{\mu}\\
\end{bmatrix}.
\label{eq:JCFwithDegeneratedEigenvalues}
\end{align}
The Jordan block is an $m\times m$ matrix when the multiplicity of the degenerate eigenvalues is $m$. 
In (\ref{eq:JCFwithDegeneratedEigenvalues}), the $\bm{\Omega}_{\mathrm{J}}$ has $m=3$. 
Note that the result of the matrix with $m=3$ can be extended to the matrix with $m>3$.
Because in the Jordan block, even if the matrix size increases, the coupling rules between adjacent oscillation modes are the same.
Then, the fundamental equation is expressed using (\ref{eq:JCFwithDegeneratedEigenvalues}) as follows
\begin{align}
\pm\ii\,\frac{\dd \, \bm{\psi}(t)}{\dd t} = \bm{\Omega}_{\mathrm{J}}\,\bm{\psi}(t).
\label{eq:EoM5}
\end{align}
where $\bm{\psi}(t)$ is the solution of the fundamental equation (\ref{eq:EoM5}).

Next, we decompose $\Omega_{\mathrm{J}}$ into a constant multiple of the unit matrix $\bm{\Omega}^{0}_{\mathrm{J}}$, and a matrix of other components $\bm{\Omega}_{\mathrm{J}}^{\mathrm{I}}$, as 
\begin{align}
\bm{\Omega}_{\mathrm{J}} = \bm{\Omega}^{0}_{\mathrm{J}}+\bm{\Omega}_{\mathrm{J}}^{\mathrm{I}}.
\label{eq:OmegaDecomp}
\end{align}
We introduce $\bm{\Omega}_\mathrm{J}^0$ and $\bm{\Omega}_\mathrm{J}^{\mathrm{I}}$ to separate $\bm{\Omega}_\mathrm{J}$ into diagonalizable matrix with well known network properties and others.
Therefore, $\bm{\Omega}_\mathrm{J}^0$ and $\bm{\Omega}_\mathrm{J}^{\mathrm{I}}$ are determined by above the choosing rule.
The fundamental equation with the network structure of linear operator (\ref{eq:OmegaDecomp}) can be expressed as follows
\begin{align}
\pm\ii\,\frac{\dd \, \bm{\psi}(t)}{\dd t} &= \left(\bm{\Omega}^{0}_{\mathrm{J}}+\bm{\Omega}_{\mathrm{J}}^{\mathrm{I}}\right)\,\bm{\psi}(t).
\label{eq:Fundamental_Omega0+OmegaI}
\end{align}
We then consider $\bm{\Psi}_0(t)$ and $\bm{\psi} _\mathrm{I}(t)$ that satisfy the following equations
\begin{align}
\pm\ii\,\frac{\dd \, \bm{\Psi}_0(t)}{\dd t} &= \bm{\Omega}^{0}_{\mathrm{J}}\,\bm{\Psi}_0(t),
\label{eq:free}\\
\pm\ii\,\frac{\dd \, \bm{\psi}_\mathrm{I}(t)}{\dd t} &= \left(\bm{\Psi}_0(-t)\,\bm{\Omega}_{\mathrm{J}}^{\mathrm{I}}\,\bm{\Psi}_0(t)\right)\,\bm{\psi}_\mathrm{I}(t),
\label{eq:T-S}
\end{align}
where $\bm{\Psi}_0(t)$ is a matrix with the initial conditions of $\bm{\Psi}_0(0) = \bm{I}$ (the $m \times m$ unit matrix) and can be expressed as follows
\begin{align}
\bm{\Psi}_0(t)=\exp[\mp\ii\, \bm{\Omega}^{0}_{\mathrm{J}}\, t].
\label{eq:Psi_0(t)}
\end{align}
In the end, $\bm{\psi}(t)$ can be written as a solution in product form as follows
\begin{align}
\bm{\psi}(t) &= \bm{\Psi}_0(t) \, \bm{\psi} _\mathrm{I}(t).
\label{eq:Psi_0psi_I}
\end{align}
Note that $\bm{\Psi}_0(t)$ is proportional to the unit matrix because the corresponding eigenvalues are degenerate.
Therefore, $\bm{\Psi}_0(t)$ can 
commute with any $m\times m$ matrix.
From the commutation relation that $\bm{\Omega}_{\mathrm{J}}\, \bm{\Psi}_0(t)=\bm{\Psi}_0(t)\, \bm{\Omega}_{\mathrm{J}}$, we obtain the following relation.
\begin{align}
\bm{\Omega}_{\mathrm{J}} &= \bm{\Psi}_0(t)\, \bm{\Omega}_{\mathrm{J}}\, \bm{\Psi}_0(-t).
\label{eq:Unitary}
\end{align}
We can write the fundamental equation (\ref{eq:Fundamental_Omega0+OmegaI}) as follows
\begin{align}
\pm\ii\, \frac{\dd \, \bm{\psi}(t)}{\dd t} &= 
\bm{\Omega}_{\mathrm{J}}\,\bm{\psi}(t) \notag\\
&=\left( \bm{\Omega}^{0}_{\mathrm{J}}+\bm{\Psi}_0(t)\, \bm{\Omega}_{\mathrm{J}}^{\mathrm{I}}\, \bm{\Psi}_0(-t) \right)\, \bm{\psi}(t).
\label{eq:H_0+H_I_Unitary}
\end{align}
This means that the fundamental equation (\ref{eq:Fundamental_Omega0+OmegaI}) is invariant under the unitary transform of $\bm{\Omega}_{\mathrm{J}}$ as follows
\[
\bm{\Omega}_{\mathrm{J}} \rightarrow \bm{\Psi}_0(t)\, \bm{\Omega}_{\mathrm{J}}\, \bm{\Psi}_0(-t), \quad 
\bm{\psi}(t) \rightarrow \bm{\Psi}_0(t)\, \bm{\psi}(t).
\]
As a result, the structure of the oscillation mode is invariant with regard to
the unitary transform given constant multiples of the unitary matrix, such as $\bm{\Psi}_0(t)$ and $\bm{\Psi}_0(-t)$.

Since $\bm{\mathcal{L}}$ is a positive semidefinite matrix, the matrix $\bm{H}$ satisfying $\bm{H}^2 = \bm{\mathcal{L}}$ is unique if and only if it is a 
positive semidefinite matrix.
Similarly, $\bm{\Omega}$ is unique for $\bm{\Lambda}$ if it is limited to a positive semidefinite matrix.
If we choose a matrix other than the positive semidefinite matrix, $\bm{\Omega}$ is not unique as $\bm{\Omega}^2 = \bm{\Lambda}$.
Since the eigenvalues of $\bm{\mathcal{L}}$ we deal with are real numbers, the real parts of the eigenvalues are non-negative.
This is guaranteed by Gerschgorin's theorem~\cite{Gerschgorin_2004}.
Now, by choosing $\bm{\Omega}$ appropriately, we try to cast both $\bm{\Lambda}$ and $\bm{\Omega}$ in Jordan canonical form. 
In the case of the Jordan block of interest, consider that the square of $\bm{\Omega}_{\mathrm{J}}$ creates a structure of Jordan canonical form.
In order to consider the necessary conditions for this, we decompose the non-diagonalizable matrix $\bm{\Omega}_{\mathrm{J}}^{\mathrm{I}}$ of $\bm{\Omega}_{\mathrm{J}}$ into the diagonal matrix $\bm{\Omega}_{\mathrm{J}}^{\mathrm{I(\mathrm{d})}}$ and the remainder matrix $\bm{\Omega}_{\mathrm{J}}^{\mathrm{I(\mathrm{a})}}$ as follows: 
\begin{align}
\bm{\Omega}_\mathrm{J}=\bm{\Omega}_\mathrm{J}^0+\bm{\Omega}_\mathrm{J}^{\mathrm{I}}
=\bm{\Omega}_\mathrm{J}^0+\bm{\Omega}_\mathrm{J}^{\mathrm{I\,(d)}}+\bm{\Omega}_\mathrm{J}^{\mathrm{I\,(a)}}.
\label{eq:OmegaJ_Decomposition}
\end{align}
The square of $\bm{\Omega}_\mathrm{J}$ is written as
\begin{align}
\bm{\Omega}_\mathrm{J}^2 &= \left(\bm{\Omega}_\mathrm{J}^0\right)^2+\left(\bm{\Omega}_\mathrm{J}^{\mathrm{I\,(d)}}\right)^2+\left(\bm{\Omega}_\mathrm{J}^{\mathrm{I\,(a)}}\right)^2
\notag\\
 &\quad +\bm{\Omega}_\mathrm{J}^0\,\bm{\Omega}_\mathrm{J}^{\mathrm{I\,(d)}} + \bm{\Omega}_\mathrm{J}^{\mathrm{I\,(d)}}\,\bm{\Omega}_\mathrm{J}^0
 \notag\\
 &\quad +\bm{\Omega}_\mathrm{J}^0\,\bm{\Omega}_\mathrm{J}^{\mathrm{I\,(a)}} + \bm{\Omega}_\mathrm{J}^{\mathrm{I\,(a)}}\,\bm{\Omega}_\mathrm{J}^0
 \notag\\
 &\quad +\bm{\Omega}_\mathrm{J}^{\mathrm{I\,(d)}}\,\bm{\Omega}_\mathrm{J}^{\mathrm{I\,(a)}}+\bm{\Omega}_\mathrm{J}^{\mathrm{I\,(a)}}\,\bm{\Omega}_\mathrm{J}^{\mathrm{I\,(d)}}.
 \label{eq:OmegaJ^2}
\end{align}
If the following anti-commutation relations 
\begin{align}
\bm{\Omega}_\mathrm{J}^0\,\bm{\Omega}_\mathrm{J}^{\mathrm{I\,(d)}} &= -\bm{\Omega}_\mathrm{J}^{\mathrm{I\,(d)}}\,\bm{\Omega}_\mathrm{J}^0,
\notag\\
\bm{\Omega}_\mathrm{J}^0\,\bm{\Omega}_\mathrm{J}^{\mathrm{I\,(a)}} &= -\bm{\Omega}_\mathrm{J}^{\mathrm{I\,(a)}}\,\bm{\Omega}_\mathrm{J}^0,
\notag\\
\bm{\Omega}_\mathrm{J}^{\mathrm{I\,(d)}}\,\bm{\Omega}_\mathrm{J}^{\mathrm{I\,(a)}} &= -\bm{\Omega}_\mathrm{J}^{\mathrm{I\,(a)}}\,\bm{\Omega}_\mathrm{J}^{\mathrm{I\,(d)}},
\notag
\end{align}
are established, all the cross-terms in (\ref{eq:OmegaJ^2}) can be canceled, yielding correspondence in Jordan canonical form as follows:  \begin{align}
\bm{\Omega}_\mathrm{J}^2 = \left(\bm{\Omega}_\mathrm{J}^0\right)^2+\left(\bm{\Omega}_\mathrm{J}^{\mathrm{I\,(d)}}\right)^2+\left(\bm{\Omega}_\mathrm{J}^{\mathrm{I\,(a)}}\right)^2.
\end{align}

In order to achieve this, according to \cite{kubo_2019}, we introduce the following Pauli matrices: 
\[
\bm{\sigma}_1=
\begin{bmatrix}
0 & 1\\
1 & 0
\end{bmatrix}, 
\quad 
\bm{\sigma}_2=
\begin{bmatrix}
0 & -\ii\\
\ii & 0
\end{bmatrix}, 
\quad 
\bm{\sigma}_3=
\begin{bmatrix}
1 & 0\\
0 & -1
\end{bmatrix}.
\]
For $i$, $j = 1$, $2$, $3$, the Pauli matrices satisfy the following relationships:
\begin{align}
& \bm{\sigma}_i^2 = \bm{E}, \notag \\
& \bm{\sigma}_i \, \bm{\sigma}_j = - \bm{\sigma}_j \, \bm{\sigma}_i ,\quad (i\not=j). \notag 
\end{align}
where $\bm{E}$ is a $2\times 2$ identity matrix.
By using the Pauli matrices, we can introduce the following matrices: 
\begin{align}
\hat{\bm{\Omega}}_{\mathrm{J}}^0 &:= \bm{\Omega}_\mathrm{J}^0\otimes\bm{\sigma}_{3}, \notag \\
\hat{\bm{\Omega}}_\mathrm{J}^{\mathrm{I}\, (\mathrm{d})} &:= \bm{\Omega}_\mathrm{J}^{\mathrm{I}\, (\mathrm{d})}\otimes\bm{\sigma}_{1}, 
\label{eq:Kronecker}\\
\hat{\bm{\Omega}}_\mathrm{J}^{\mathrm{I}\, (\mathrm{a})} &:= \bm{\Omega}_\mathrm{J}^{\mathrm{I}\, (\mathrm{a})}\otimes\bm{\sigma}_{2}, \notag
\end{align}
where $\otimes$ represents the Kronecker product.
By extending the matrix using the Pauli matrices, the square of $\hat{\bm{\Omega}}_\mathrm{J}$ can be expressed as 
\begin{align}
\hat{\bm{\Omega}}_\mathrm{J}^2 = \left(\hat{\bm{\Omega}}_\mathrm{J}^0\right)^2+\left(\hat{\bm{\Omega}}_\mathrm{J}^{\mathrm{I\,(d)}}\right)^2+\left(\hat{\bm{\Omega}}_\mathrm{J}^{\mathrm{I\,(a)}}\right)^2.
\label{eq:HatOmegaJ^2}
\end{align}

By using the Pauli matrices, we can rewrite the fundamental equation (\ref{eq:Fundamental_Omega0+OmegaI}) as follows: 
\begin{align}
\pm\mathrm{i}\,\frac{\mathrm{d}\, \hat{\bm{\psi}}(t)}{\mathrm{d}t} = \hat{\bm{\Omega}}_\mathrm{J}\,\hat{\bm{\psi}}(t) =\left(\hat{\bm{\Omega}}_\mathrm{J}^0+
\hat{\bm{\Omega}}_\mathrm{J}^{\mathrm{I}\, (\mathrm{d})}+\hat{\bm{\Omega}}_\mathrm{J}^{\mathrm{I}\, (\mathrm{a})}\right)\,\hat{\bm{\psi}}(t),
\label{eq:HatPsiH} 
\end{align}
where $\hat{\bm{\Omega}}_\mathrm{J}$ is a $2m \times 2m$ square matrix.
$\hat{\bm{\psi}}(t)$ is a $2m$-dimensional vector defined as 
\begin{align}
\hat{\bm{\psi}}(t) = {}^{t}\!(\psi_{1}^{\uparrow}(t),\,\psi_{1}^{\downarrow}(t),\,\psi_{2}^{\uparrow}(t),\,\psi_{2}^{\downarrow}(t), \ldots,\psi_{m}^{\uparrow}(t),\,\psi_{m}^{\downarrow}(t)).
\end{align}
Note that the extension of states $\psi_{}^{\uparrow}(t)$ and $\psi_{}^{\downarrow}(t)$ is an algebraic technique to simplify the computation. 

Next, we introduce the unitary-transformed fundamental equation, 
\begin{align} 
&\pm\ii\,\frac{\dd \, \hat{\bm{\psi}}(t)}{\dd t} 
\notag\\
&= \left(\hat{\bm{\Omega}}_{\mathrm{J}}^0 + \e^{\pm\ii\, \hat{\bm{\Omega}}_{\mathrm{J}}^0\, t}\left( \hat{\bm{\Omega}}_\mathrm{J}^{\mathrm{I}\, (\mathrm{d})} + \hat{\bm{\Omega}}_\mathrm{J}^{\mathrm{I}\, (\mathrm{a})} \right)\,\e^{\mp\ii\, \hat{\bm{\Omega}}_{\mathrm{J}}^0\, t} \right)\, \hat{\bm{\psi}}(t).
\label{eq:FundamentalUnitaryOmegaJ}
\end{align}
as shown in~\cite{kubo_2019}.
The fundamental equation (\ref{eq:EoM5}) for $m$ dimensional state vectors is invariant with respect to the unitary transform.
However, the fundamental equation (\ref{eq:HatPsiH}) for the $2m$-dimensional state vector is not invariant under unitary transformation since the fundamental equation (\ref{eq:FundamentalUnitaryOmegaJ}) is different from (\ref{eq:HatPsiH}).
Since the fundamental equation describing the actual user dynamics is the equation (\ref{eq:HatPsiH}) before unitary transformation, the solution of equation (\ref{eq:FundamentalUnitaryOmegaJ}) after the unitary transform does not necessarily represent the actual user dynamics.
However, since the unitary-transformed fundamental equation has an interesting structure, our previous study ~\cite{kubo_2019} investigated the nature of the user dynamics described by the unitary-transformed equation and has revealed interesting properties that appear in it.

\subsection{Flaming Phenomena Caused by Degenerate Eigenvalues}
One of the facts revealed
in \cite{kubo_2019} is that the unitary-transformed fundamental equation of networks (\ref{eq:FundamentalUnitaryOmegaJ}) describes divergence corresponding to the flaming phenomena in spite of the fact that all the eigenvalues are real numbers.
This subsection briefly explains its characteristics.

\begin{figure}[b]
\begin{center}
\includegraphics[width=0.8\linewidth]{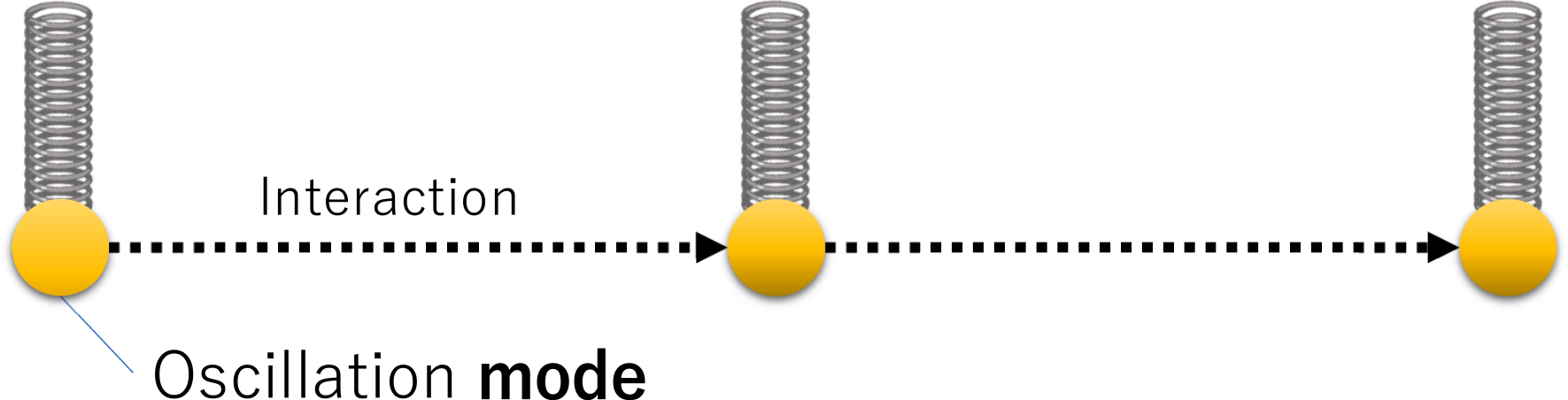}
\end{center}
\caption{Coupling states of 3 oscillation modes}
\label{fig:3OscillationModes}
\end{figure}

Figure \ref{fig:3OscillationModes} shows the state of three oscillation modes coupled as an example of (\ref{eq:FundamentalUnitaryOmegaJ}).
If the degenerate eigenvalue in (\ref{eq:FundamentalUnitaryOmegaJ}) is $\omega$, then $\hat{\bm{\Omega}}_{\mathrm{J}}^0,\,\hat{\bm{\Omega}}_\mathrm{J}^{\mathrm{I}\, (\mathrm{d})},\, \hat{\bm{\Omega}}_\mathrm{J}^{\mathrm{I}\, (\mathrm{a})}$ can be expressed as follows
\begin{align}
\hat{\bm{\Omega}}_{\mathrm{J}}^0 &= \bm{\Omega}_\mathrm{J}^0\otimes\bm{\sigma}_{3}=
\scalebox{1.0}{$\displaystyle
\begin{bmatrix}
\omega & 0 & 0\\
0 & \omega & 0\\
0 & 0 & \omega
\end{bmatrix} $}\otimes\bm{\sigma}_{3}, \notag \\ 
\hat{\bm{\Omega}}_\mathrm{J}^{\mathrm{I}\, (\mathrm{d})} &= \bm{\Omega}_\mathrm{J}^{\mathrm{I}\, (\mathrm{d})}\otimes\bm{\sigma}_{1}=
\scalebox{1.0}{$\displaystyle
\begin{bmatrix}
d & 0 & 0\\
0 & d & 0\\
0 & 0 & d
\end{bmatrix} $}\otimes\bm{\sigma}_{1}, 
\label{eq:EachOmega} \\
\hat{\bm{\Omega}}_\mathrm{J}^{\mathrm{I}\, (\mathrm{a})} &= \bm{\Omega}_\mathrm{J}^{\mathrm{I}\, (\mathrm{a})}\otimes\bm{\sigma}_{2}=
\scalebox{1.0}{$\displaystyle
\begin{bmatrix}
0 & 1 & 0\\
0 & 0 & 1\\
0 & 0 & 0
\end{bmatrix} $}\otimes\bm{\sigma}_{2},
\notag
\end{align}
where $\omega+d$ is the eigenfrequency of the degenerate oscillation mode.

\begin{figure}[b]
\begin{center}
\includegraphics[width=0.8\linewidth]{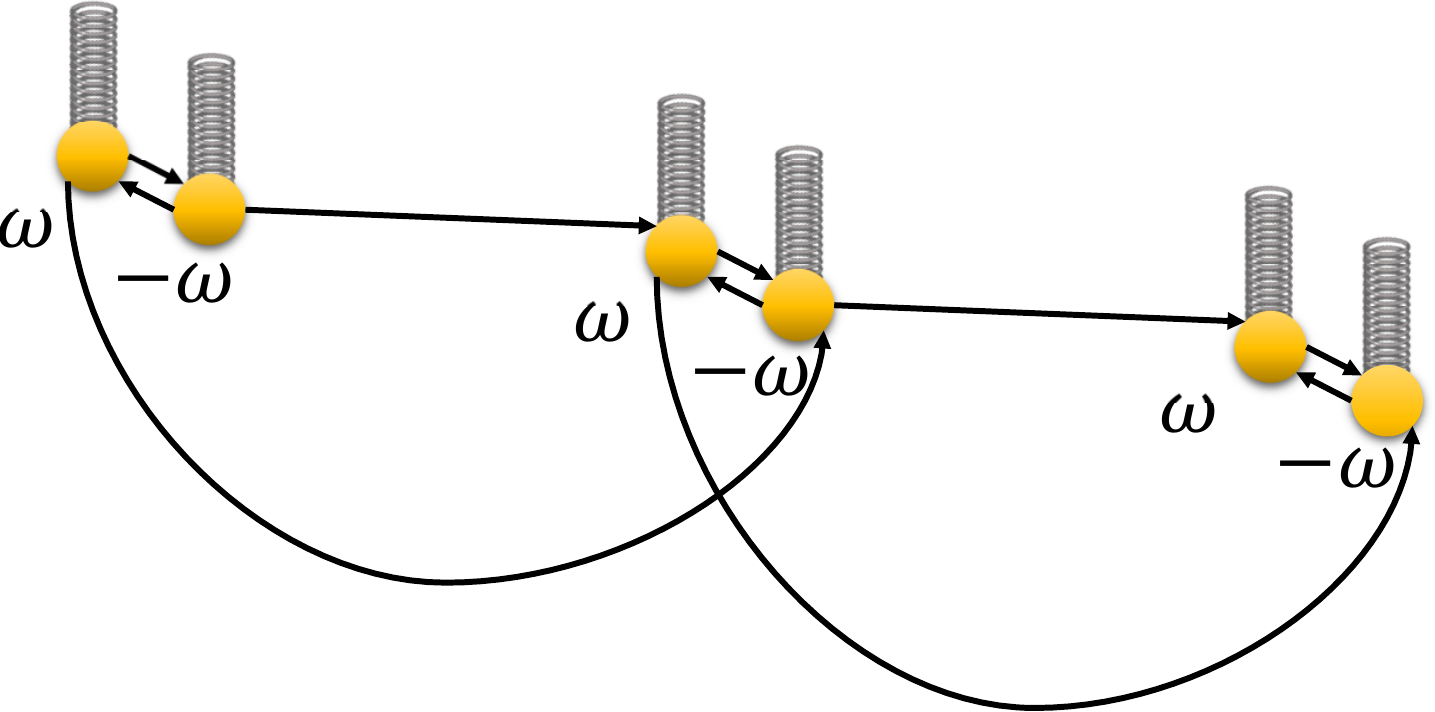}
\end{center}
\caption{Coupling states of 6 oscillation modes}
\label{fig:6OscillationModes}
\end{figure}

From (\ref{eq:EachOmega}), the linear operator in the fundamental equation (\ref{eq:FundamentalUnitaryOmegaJ}) can be written as follows
\begin{align}
&\hat{\bm{\Omega}}_{\mathrm{J}}^0+\e^{+\ii \, \hat{\bm{\Omega}}_{\mathrm{J}}^0\, t}\left(\hat{\bm{\Omega}}_\mathrm{J}^{\mathrm{I}\, (\mathrm{d})}+\hat{\bm{\Omega}}_\mathrm{J}^{\mathrm{I}\, (\mathrm{a})}\right)\, \e^{-\ii \, \hat{\bm{\Omega}}_{\mathrm{J}}^0\, t} \notag \\
& =\bm{\Omega}_\mathrm{J}^0\otimes\, \bm{\sigma}_{3} 
\notag\\
&\qquad {}+\e^{+\ii \, \bm{\Omega}_\mathrm{J}^0\otimes\, \bm{\sigma}_{3}\, t}\left(\bm{\Omega}_\mathrm{J}^{\mathrm{I}\, (\mathrm{d})}\otimes\bm{\sigma}_{1}
+\bm{\Omega}_\mathrm{J}^{\mathrm{I}\, (\mathrm{a})}\otimes\bm{\sigma}_{2}\right)\, \e^{-\ii \, \bm{\Omega}_\mathrm{J}^0\otimes\, \bm{\sigma}_{3}\, t} \notag \\
& = \scalebox{0.73}{$\displaystyle
\left[
	\begin{array}{cccccc}
	\omega & d \, \e^{-\ii 2\,\omega\, t} & 0 & -\ii \, \e^{-\ii 2\,\omega\, t} & 0 & 0 \\
	d \, \e^{+\ii 2\,\omega\, t} & -\omega & \ii \, \e^{+\ii 2\,\omega\, t} & 0 & 0 & 0 \\
	0 & 0 & \omega & d \, \e^{-\ii 2\,\omega\, t} & 0 &  -\ii \, \e^{-\ii 2\,\omega\, t} \\
	0 & 0 & d \, \e^{+\ii 2\,\omega\, t} & -\omega & \ii \, \e^{+\ii 2\,\omega\, t} & 0 \\
	0 & 0 & 0 & 0 & \omega & d \, \e^{-\ii 2\,\omega\, t} \\
	0 & 0 & 0 & 0 & d\, \e^{+\ii 2\,\omega\, t} & -\omega
	\end{array}
  \right] $}.
  \label{eq:OmegaJ_PauliUnitary} 
\end{align}
Then, the fundamental equation (\ref{eq:FundamentalUnitaryOmegaJ}) is written as follows
\begin{align}
& \ii \, \frac{\mathrm{d}}{\mathrm{d}t}\, 
 \scalebox{1.0}{$\displaystyle
 \hat{\bm{\psi}}(t) $} 
\notag\\
&= \scalebox{0.7}{$\displaystyle
\left[
	\begin{array}{cccccc}
	\omega & d \, \e^{-\ii 2\,\omega\, t} & 0 & -\ii \, \e^{-\ii 2\,\omega\, t} & 0 & 0 \\
	d \, \e^{+\ii 2\,\omega\, t} & -\omega & \ii \, \e^{+\ii 2\,\omega\, t} & 0 & 0 & 0 \\
	0 & 0 & \omega & d \, \e^{-\ii 2\,\omega\, t} & 0 &  -\ii \, \e^{-\ii 2\,\omega\, t} \\
	0 & 0 & d \, \e^{+\ii 2\,\omega\, t} & -\omega & \ii \, \e^{+\ii 2\,\omega\, t} & 0 \\
	0 & 0 & 0 & 0 & \omega & d \, \e^{-\ii 2\,\omega\, t} \\
	0 & 0 & 0 & 0 & d\, \e^{+\ii 2\,\omega\, t} & -\omega
	\end{array}
  \right] $} \, \scalebox{1.0}{$\displaystyle
  \hat{\bm{\psi}}(t). $}
  \label{eq:Fundamental_PauliUnitaryOmega}
\end{align}
The coupled state of the oscillation mode corresponding to (\ref{eq:Fundamental_PauliUnitaryOmega}) is shown in Fig.~\ref{fig:6OscillationModes}.

Let us consider the solution of the degenerate modes of the fundamental equation (\ref{eq:Fundamental_PauliUnitaryOmega}).
First, we introduce several functions $\delta_{1}^{\uparrow}(t)$, $\delta_{1}^{\downarrow}(t)$, $\delta_{2}^{\uparrow}(t)$, $\delta_{2}^{\downarrow}(t)$, $\delta_{3}^{\uparrow}(t)$, $\delta_{3}^{\downarrow}(t)$ that represent the deviation from the eigenfrequency $\mp\omega$ of the symmetrizable fundamental equations.
The solutions to the fundamental equations take the following form
\begin{align}
\psi_{1}^{\uparrow}(t) &= \e^{-\ii\theta_{1}^{\uparrow}(t)} = \e^{-\ii\omega\,t+\ii\delta_{1}^{\uparrow}(t)}, 
\label{eq:PsiSolution1}\\
\psi_{1}^{\downarrow}(t) &= \e^{+\ii\theta_{1}^{\downarrow}(t)} = \e^{+\ii\omega\,t+\ii\delta_{1}^{\downarrow}(t)}.
\label{eq:PsiSolution2}\\
\psi_{2}^{\uparrow}(t) &= \e^{-\ii\theta_{2}^{\uparrow}(t)} = \e^{-\ii\omega\,t+\ii\delta_{2}^{\uparrow}(t)}, 
\label{eq:PsiSolution3}\\
\psi_{2}^{\downarrow}(t) &= \e^{+\ii\theta_{2}^{\downarrow}(t)} = \e^{+\ii\omega\,t+\ii\delta_{2}^{\downarrow}(t)}.
\label{eq:PsiSolution4}\\
\psi_{3}^{\uparrow}(t) &= \e^{-\ii\theta_{3}^{\uparrow}(t)} = \e^{-\ii\omega\,t+\ii\delta_{3}^{\uparrow}(t)}, 
\label{eq:PsiSolution5}\\
\psi_{3}^{\downarrow}(t) &= \e^{+\ii\theta_{3}^{\downarrow}(t)} = \e^{+\ii\omega\,t+\ii\delta_{3}^{\downarrow}(t)}.
\label{eq:PsiSolution6}
\end{align}
Substituting (\ref{eq:PsiSolution1})--(\ref{eq:PsiSolution6}) into the fundamental equation (\ref{eq:Fundamental_PauliUnitaryOmega}), we obtain the time evolution equations for the phases.
Taking the time evolution equation for $\delta_{1}^{\uparrow}(t)$ and $\delta_{1}^{\downarrow}(t)$ as an example, we get
\begin{align}
\frac{\dd \, \delta_{1}^{\uparrow}(t)}{\dd t} 
=-d\, \e^{+\ii \, (\delta_{1}^{\downarrow}(t)-\delta_{1}^{\uparrow}(t))}
+\ii \, \e^{+\ii \, (\delta_{2}^{\downarrow}(t)-\delta_{1}^{\uparrow}(t))},
\label{eq:ddelta1}\\
\frac{\dd \, \delta_{1}^{\downarrow}(t)}{\dd t} 
=-d\, \e^{+\ii \, (\delta_{1}^{\uparrow}(t)-\delta_{1}^{\downarrow}(t))}
-\ii \, \e^{+\ii \, (\delta_{2}^{\uparrow}(t)-\delta_{1}^{\downarrow}(t))}.
\label{eq:ddelta2}
\end{align}
Because the phase difference is not always a real number, $\delta_{1}^{\uparrow}(t)$ and $\delta_{1}^{\downarrow}(t)$ are written as complex numbers as follows: 
\begin{align}
\delta_{1}^{\uparrow}(t)&=\mathrm{Re}[\delta_{1}^{\uparrow}(t)] +\ii \, \mathrm{Im}[\delta_{1}^{\uparrow}(t)], 
\label{eq:DefPhaseDifference}\\
\delta_{1}^{\downarrow}(t)&=\mathrm{Re}[\delta_{1}^{\downarrow}(t)] + \ii \, \mathrm{Im}[\delta_{1}^{\downarrow}(t)].
\label{eq:DefPhaseDifference2}
\end{align}
Substituting (\ref{eq:DefPhaseDifference}) and (\ref{eq:DefPhaseDifference2}) into (\ref{eq:ddelta1}) and (\ref{eq:ddelta2}) gives the following time evolution equations for the real and imaginary parts of the phase difference.
\begin{align}
&\frac{\dd \, \mathrm{Re}[\delta_{1}^{\uparrow}(t)]}{\dd t}= 
\notag \\ 
&+d \, \e^{-\mathrm{Im}[\delta_{1}^{\downarrow}(t)]+\mathrm{Im}[\delta_{1}^{\uparrow}(t)]} \, 
\sin\left(\mathrm{Re}[\delta_{1}^{\downarrow}(t)]-\mathrm{Re}[\delta_{1}^{\uparrow}(t)]-\frac{\pi}{2}\right)
\notag \\
&+\e^{-\mathrm{Im}[\delta_{2}^{\downarrow}(t)]+\mathrm{Im}[\delta_{1}^{\uparrow}(t)]} \, 
\sin\left(\mathrm{Re}[\delta_{2}^{\downarrow}(t)]-\mathrm{Re}[\delta_{1}^{\uparrow}(t)]-\pi\right),
\label{eq:Red1u}\\
&\frac{\dd \, \mathrm{Re}[\delta_{1}^{\downarrow}(t)]}{\dd t}= 
\notag \\ 
&+d \, \e^{+\mathrm{Im}[\delta_{1}^{\downarrow}(t)]-\mathrm{Im}[\delta_{1}^{\uparrow}(t)]} \, 
\sin\left(\mathrm{Re}[\delta_{1}^{\uparrow}(t)]-\mathrm{Re}[\delta_{1}^{\downarrow}(t)]- \frac{\pi}{2}\right)
\notag \\
&+\e^{+\mathrm{Im}[\delta_{1}^{\downarrow}(t)]-\mathrm{Im}[\delta_{2}^{\uparrow}(t)]} \, 
\sin\left(\mathrm{Re}[\delta_{2}^{\uparrow}(t)]-\mathrm{Re}[\delta_{1}^{\downarrow}(t)]\right),
\label{eq:Red1d}\\
&\frac{\dd \, \mathrm{Im}[\delta_{1}^{\uparrow}(t)]}{\dd t}= 
\notag \\ 
&+d \, \e^{-\mathrm{Im}[\delta_{1}^{\downarrow}(t)]+\mathrm{Im}[\delta_{1}^{\uparrow}(t)]} \, 
\sin\left(\mathrm{Re}[\delta_{1}^{\downarrow}(t)]-\mathrm{Re}[\delta_{1}^{\uparrow}(t)]+\pi\right)
\notag \\
&+\e^{-\mathrm{Im}[\delta_{2}^{\downarrow}(t)]+\mathrm{Im}[\delta_{1}^{\uparrow}(t)]} \, 
\sin\left(\mathrm{Re}[\delta_{2}^{\downarrow}(t)]-\mathrm{Re}[\delta_{1}^{\uparrow}(t)]+\frac{\pi}{2}\right),
\label{eq:Imd1u}\\
&\frac{\dd \, \mathrm{Im}[\delta_{1}^{\downarrow}(t)]}{\dd t}= 
\notag \\ 
&+d \, \e^{+\mathrm{Im}[\delta_{1}^{\downarrow}(t)]-\mathrm{Im}[\delta_{1}^{\uparrow}(t)]} \, 
\sin\left(\mathrm{Re}[\delta_{1}^{\uparrow}(t)]-\mathrm{Re}[\delta_{1}^{\downarrow}(t)]-\pi\right)
\notag \\
&+\e^{+\mathrm{Im}[\delta_{1}^{\downarrow}(t)]-\mathrm{Im}[\delta_{2}^{\uparrow}(t)]} \, 
\sin\left(\mathrm{Re}[\delta_{2}^{\uparrow}(t)]-\mathrm{Re}[\delta_{1}^{\downarrow}(t)]-\frac{\pi}{2}\right).
\label{eq:Imd1d}
\end{align}
The time evolution equations (\ref{eq:Red1u}) and (\ref{eq:Red1d}) of the real parts of the phase difference are similar to the model of phase synchronization called the Kuramoto model~\cite{kuramoto} or the Sakaguchi model~\cite{sakaguchi_1986}.
We call these synchronization models in this paper the Kuramoto model.
The significant difference from the Kuramoto model lies in the coefficients of the sine function in (\ref{eq:Red1u}) and (\ref{eq:Red1d}). 
In our model, the coefficients of the sine function are not constant.
However, when the exponential function, which is a coefficient of the sine function, is considered to be sufficiently large, phase synchronization occurs, as in the Kuramoto model, where the argument of the sine function becomes zero.
Since there are two or more different terms of the sinusoidal function, such as (\ref{eq:Red1u}) and (\ref{eq:Red1d}), we should confirm if the corresponding phase synchronizations are compatible or not. 
Since phase synchronization is incompatible only if one of the terms becomes dominant, Kuramoto-model-like phase synchronization occurs and the phase difference is fixed at a specific value.

On the other hand, the time evolution equations (\ref{eq:Imd1u}) and (\ref{eq:Imd1d}) for the imaginary part do not trigger phase synchronization, but they are influenced by the difference of real parts of phases. 
If some of the phase differences of real parts of phases are fixed, the sign of the sine functions of the first and second terms on the right-hand side are fixed.
Then, the rate of change of the imaginary part of the phase becomes fixed at either positive or negative values.

Substituting (\ref{eq:DefPhaseDifference}) into (\ref{eq:PsiSolution1}) and (\ref{eq:PsiSolution2}), $\psi_{1}^{\uparrow}(t)$ and $\psi_{1}^{\downarrow}(t)$ are written as follows
\begin{align}
\psi_{1}^{\uparrow}(t) &= \e^{-\mathrm{Im}[\delta_{1}^{\uparrow}(t)]}\,\e^{-\mathrm{i}\, \omega\,t+\mathrm{i}\, \mathrm{Re}[\delta_{1}^{\uparrow}(t)]}, \label{eq:PsiDeltaSol1}\\
\psi_{1}^{\downarrow}(t) &= \e^{-\mathrm{Im}[\delta_{1}^{\downarrow}(t)]}\,\e^{+\mathrm{i}\, \omega\,t+\mathrm{i}\, \mathrm{Re}[\delta_{1}^{\downarrow}(t)]}.
\label{eq:PsiDeltaSol2}
\end{align}
In (\ref{eq:PsiDeltaSol1}), (\ref{eq:PsiDeltaSol2}), as the phase synchronization of $\mathrm{Re}[\delta_{1}^{\uparrow}(t)]$, $\mathrm{Re}[\delta_{1}^{\downarrow}(t)]$ progresses, $\e^{-\mathrm{Im}[\delta_{1}^{\uparrow}(t)]}$, $\e^{-\mathrm{Im}[\delta_{1}^{\downarrow}(t)]}$ increases exponentially with time.
Therefore if $\mathrm{Im}[\delta_{1}^{\uparrow}(t)]$ or $\mathrm{Im}[\delta_{1}^{\downarrow}(t)]$ approaches $-\infty$, the amplitude of $\psi_{1}^{\uparrow}(t)$ or $\psi_{1}^{\downarrow}(t)$ diverges. This means oscillation energy divergence, which is one model that can describe flaming phenomena.
This result shows that divergence might occur even if eigenvalues are real numbers given that the eigenvalues are degenerate.

\subsection{Fundamental Equation of User Dynamics with Degeneration of Eigenvalue $0$}
As mentioned above, the unitary-transformed fundamental equation (\ref{eq:FundamentalUnitaryOmegaJ}) does not necessarily represent the dynamics of actual networks, and therefore might not lead directly to flaming phenomena in real networks.
The dynamics of actual networks are described by the fundamental equation (\ref{eq:HatPsiH}), to which the unitary transformation is not applied.
The extended fundamental equation (\ref{eq:HatPsiH}) for  the network model shown in Fig.~\ref{fig:3OscillationModes} is written as follows:
\begin{align}
\ii \, \frac{\mathrm{d}}{\mathrm{d}t}\,
 \scalebox{1.0}{$\displaystyle
 \hat{\bm{\psi}}(t) $}
= \scalebox{0.7}{$\displaystyle
\left[
	\begin{array}{cccccc}
	\omega & d & 0 & -\ii & 0 & 0 \\
	d & -\omega & \ii & 0 & 0 & 0 \\
	0 & 0 & \omega & d & 0 &  -\ii \\
	0 & 0 & d & -\omega & \ii & 0 \\
	0 & 0 & 0 & 0 & \omega & d \\
	0 & 0 & 0 & 0 & d & -\omega
	\end{array}
  \right] $}\, \scalebox{1.0}{$\displaystyle
  \hat{\bm{\psi}}(t). $}
  \label{eq:Fundamental_PauliOmega}
\end{align}
If we cast the solutions to the equation as (\ref{eq:PsiSolution1})--(\ref{eq:PsiSolution6}), and then enter them into (\ref{eq:Fundamental_PauliOmega}), we can derive the time evolution equations for the phases by a procedure similar to that  shown in the previous subsection. 
The time evolution equations for $\delta_{1}^{\uparrow}(t)$ and $\delta_{1}^{\downarrow}(t)$ are written as follows:
\begin{align}
&\frac{\dd \, \mathrm{Re}[\delta_{1}^{\uparrow}(t)]}{\dd t} 
\notag\\
&= +d\, \e^{-\mathrm{Im}[\delta_{1}^{\downarrow}(t)]+\mathrm{Im}[\delta_{1}^{\uparrow}(t)]} 
\notag\\
&\qquad\times \Big{[} \cos (2\, \omega\, t)\, \sin \left(\mathrm{Re}[\delta_{1}^{\downarrow}(t)]-\pi/2-\mathrm{Re}[\delta_{1}^{\uparrow}(t)]\right)
\notag \\
&\qquad\qquad\qquad {} + \sin  (2\, \omega\, t)\, \sin \left(\mathrm{Re}[\delta_{1}^{\downarrow}(t)]-\mathrm{Re}[\delta_{1}^{\uparrow}(t)]\right)\Big{]} 
\notag \\ 
&\quad {} + \e^{-\mathrm{Im}[\delta_{2}^{\downarrow}(t)]+\mathrm{Im}[\delta_{1}^{\uparrow}(t)]} 
\notag\\
&\qquad\times \Big{[} \cos (2\, \omega\, t)\, \sin \left(\mathrm{Re}[\delta_{2}^{\downarrow}(t)]+\pi-\mathrm{Re}[\delta_{1}^{\uparrow}(t)]\right)
\notag \\
&\qquad\qquad {}  + \sin  (2\, \omega\, t)\, \sin \left(\mathrm{Re}[\delta_{2}^{\downarrow}(t)]-\pi/2-\mathrm{Re}[\delta_{1}^{\uparrow}(t)]\right)\Big{]},
\label{eq:Red1u_NonUnitary}
\end{align}
\begin{align}
&\frac{\dd \, \mathrm{Re}[\delta_{1}^{\downarrow}(t)]}{\dd t} 
\notag\\
&= +d\, \e^{-\mathrm{Im}[\delta_{1}^{\uparrow}(t)]+\mathrm{Im}[\delta_{1}^{\downarrow}(t)]}
\notag\\
&\qquad\times \Big{[} \cos (2\, \omega\, t)\, \sin \left(\mathrm{Re}[\delta_{1}^{\uparrow}(t)]-\pi/2-\mathrm{Re}[\delta_{1}^{\downarrow}(t)]\right)
\notag \\
&\qquad\qquad {} + \sin  (2\, \omega\, t)\, \sin \left(\mathrm{Re}[\delta_{1}^{\uparrow}(t)]+\pi-\mathrm{Re}[\delta_{1}^{\downarrow}(t)]\right)\Big{]} 
\notag \\ 
&\quad {} + \e^{-\mathrm{Im}[\delta_{2}^{\uparrow}(t)]+\mathrm{Im}[\delta_{1}^{\downarrow}(t)]} 
\notag\\
&\qquad\times \Big{[} \cos (2\, \omega\, t)\, \sin \left(\mathrm{Re}[\delta_{2}^{\uparrow}(t)]-\mathrm{Re}[\delta_{1}^{\downarrow}(t)]\right)
\notag \\
&\qquad\qquad {} + \sin  (2\, \omega\, t)\, \cos \left(\mathrm{Re}[\delta_{2}^{\uparrow}(t)]-\pi/2-\mathrm{Re}[\delta_{1}^{\downarrow}(t)]\right)\Big{]},
\label{eq:Red1d_NonUnitary}
\end{align}
\begin{align}
&\frac{\dd \, \mathrm{Im}[\delta_{1}^{\uparrow}(t)]}{\dd t} 
\notag\\
&= +d\, \e^{-\mathrm{Im}[\delta_{1}^{\downarrow}(t)]+\mathrm{Im}[\delta_{1}^{\uparrow}(t)]} 
\notag\\
&\qquad\times \Big{[} \cos (2\, \omega\, t)\, \sin \left(\mathrm{Re}[\delta_{1}^{\downarrow}(t)]+\pi-\mathrm{Re}[\delta_{1}^{\uparrow}(t)]\right)
\notag \\
&\qquad\qquad {} + \sin  (2\, \omega\, t)\, \sin \left(\mathrm{Re}[\delta_{1}^{\downarrow}(t)]-\pi/2-\mathrm{Re}[\delta_{1}^{\uparrow}(t)]\right)\Big{]} 
\notag \\ 
&\quad {} + \e^{-\mathrm{Im}[\delta_{2}^{\downarrow}(t)]+\mathrm{Im}[\delta_{1}^{\uparrow}(t)]} 
\notag\\
&\qquad\times \Big{[} \cos (2\, \omega\, t)\, \sin \left(\mathrm{Re}[\delta_{2}^{\downarrow}(t)]+\pi/2-\mathrm{Re}[\delta_{1}^{\uparrow}(t)]\right)
\notag \\
&\qquad\qquad\quad {} + \sin  (2\, \omega\, t)\, \sin \left(\mathrm{Re}[\delta_{2}^{\downarrow}(t)]+\pi-\mathrm{Re}[\delta_{1}^{\uparrow}(t)]\right)\Big{]},
\label{eq:Imd1u_NonUnitary}
\end{align}
\begin{align}
&\frac{\dd \, \mathrm{Im}[\delta_{1}^{\downarrow}(t)]}{\dd t} 
\notag\\
&= +d\, \e^{-\mathrm{Im}[\delta_{1}^{\uparrow}(t)]+\mathrm{Im}[\delta_{1}^{\downarrow}(t)]} 
\notag\\
&\quad\times \Big{[} \cos (2\, \omega\, t)\, \sin \left(\mathrm{Re}[\delta_{1}^{\uparrow}(t)]+\pi-\mathrm{Re}[\delta_{1}^{\downarrow}(t)]\right)
\notag \\
&\qquad\qquad {} + \sin  (2\, \omega\, t)\, \cos \left(\mathrm{Re}[\delta_{1}^{\uparrow}(t)]+\pi/2-\mathrm{Re}[\delta_{1}^{\downarrow}(t)]\right)\Big{]} 
\notag \\ 
&\quad {} + \e^{-\mathrm{Im}[\delta_{2}^{\uparrow}(t)]+\mathrm{Im}[\delta_{1}^{\downarrow}(t)]} 
\notag\\
&\quad\times \Big{[} \cos (2\, \omega\, t)\, \sin \left(\mathrm{Re}[\delta_{2}^{\uparrow}(t)]-\pi/2-\mathrm{Re}[\delta_{1}^{\downarrow}(t)]\right)
\notag \\
&\qquad\qquad {} + \sin  (2\, \omega\, t)\, \sin \left(\mathrm{Re}[\delta_{2}^{\uparrow}(t)]+\pi-\mathrm{Re}[\delta_{1}^{\downarrow}(t)]\right)\Big{]}.
\label{eq:Imd1d_NonUnitary}
\end{align}
In (\ref{eq:Red1u_NonUnitary}),--(\ref{eq:Imd1d_NonUnitary}), unlike (\ref{eq:Red1u})--(\ref{eq:Imd1d}), the coefficients of $\sin(2\, \omega\, t)$ or $\cos(2\, \omega\, t)$ appear in each term, making it difficult for Kuramoto-model-like phase synchronization to occur.

\begin{figure*}
\begin{center}
\includegraphics[width=0.7\linewidth]{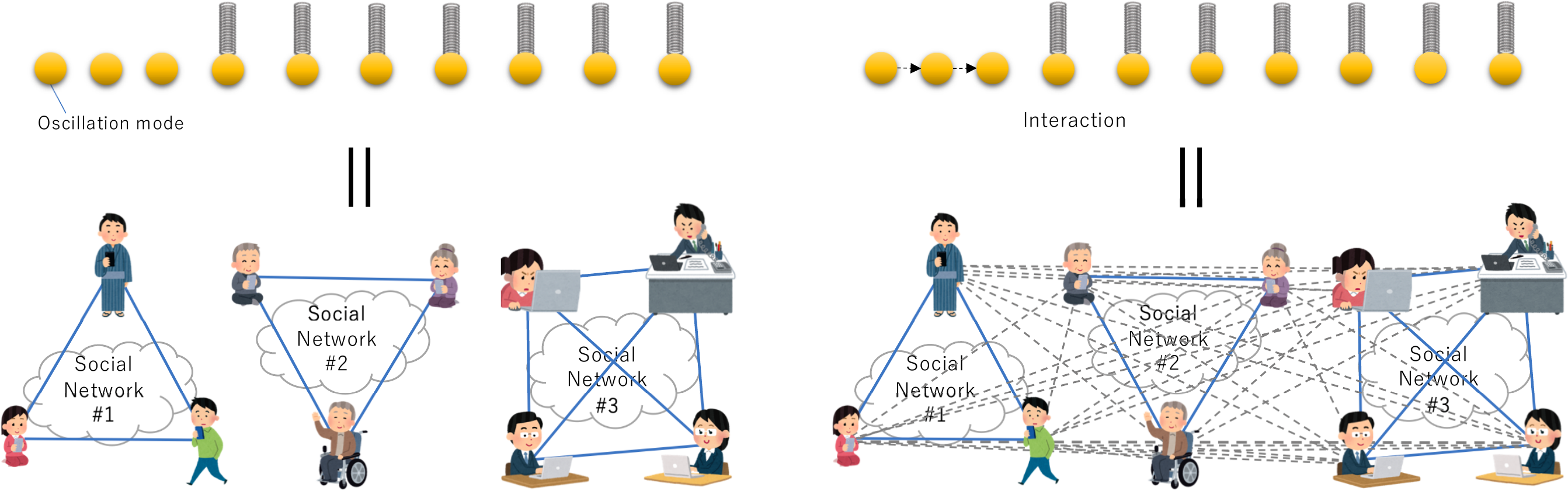}
\end{center}
\caption{Network structure when oscillation mode $0$ instances are coupled}
\label{fig:MSN}
\end{figure*}

However, there is an exceptional situation wherein the unitary-transformed fundamental equation (\ref{eq:Fundamental_PauliUnitaryOmega}) of the networks matches the fundamental equation (\ref{eq:Fundamental_PauliOmega}) of the real network.
It is when the degenerate eigenvalue is zero as follows:
\begin{align}
\ii \, \frac{\mathrm{d}}{\mathrm{d}t}\,
 \scalebox{1.0}{$\displaystyle
 \hat{\bm{\psi}}(t) $}
= \scalebox{0.7}{$\displaystyle
\left[
	\begin{array}{cccccc}
	0 & d & 0 & -\ii & 0 & 0 \\
	d & 0 & \ii & 0 & 0 & 0 \\
	0 & 0 & 0 & d & 0 &  -\ii \\
	0 & 0 & d & 0 & \ii & 0 \\
	0 & 0 & 0 & 0 & 0 & d \\
	0 & 0 & 0 & 0 & d & 0
	\end{array}
  \right] $} \, \scalebox{1.0}{$\displaystyle
  \hat{\bm{\psi}}(t). $}
  \label{eq:Fundamental_PauliOmega_ZeroOmega}
\end{align}
This is equivalent to the case of $\omega=0$ in the unitary-transformed fundamental equation (\ref{eq:Fundamental_PauliUnitaryOmega}), which suggests that flaming phenomena may occur.

Note that the matrix that appeared in (\ref{eq:Fundamental_PauliOmega_ZeroOmega}) has only one eigenvalue of $0$. 
The matrix appears to be similar in the Jordan canonical form, but it is not. 
The non-diagonal component causes coupling of the separated subnetworks by weak ties (see Appendix).
Thus, the situation where eigenvalue $0$ couples as a Jordan block corresponds to the situation where coupling occurs in the divided communities shown on the left of Fig.~\ref{fig:MSN} which resolves the network fragmentation, as shown on the right of Fig.~\ref{fig:MSN}.
Since we are considering a limited change in which only the oscillation modes with eigenvalues of $0$ couple; this represents the situation in which the divided communities are weakly coupled with each other.

\section{Numerical Experiments}
\subsection{Outline of Numerical Experiments}
The previous section considered the fundamental equations of user dynamics with degenerate eigenvalues and showed that flaming phenomena might occur even when all eigenvalues are real numbers if eigenvalue degeneration occurs with eigenvalue $0$.
However, since several parameters are involved in the occurrence of flaming phenomena, we have to confirm whether flaming phenomena can actually occur.
In this section, we use numerical experiments to confirm that the flaming phenomena associated with the degeneration of eigenvalue $0$ can actually occur. 

The target of the numerical experiments is the system described by the fundamental equation (\ref{eq:Fundamental_PauliOmega_ZeroOmega}). 
This is the situation described in Sec.~3.2, where the three divided communities shown in Fig.~3 have weak ties and coupling between the three zero eigenvalues, represented by the Jordan canonical form in appearance, has occurred.
Given the initial conditions shown in Table~\ref{table:parameter}, we evaluate the time evolution of the phase of the solution.
The time evolution of the phase of the solution is analyzed separately for real and imaginary parts, and each time evolution equation is evaluated from the following perspectives. 
\begin{itemize}
\item For the time evolution of the real part of the phase, we investigate whether the phase difference is fixed due to Kuramoto-like synchronization or not.
If the phase difference is not fixed, we also investigate whether the phase difference is limited to a specific range.
\item For the time evolution of the imaginary part of the phase, we investigate the influence of the behavior of the phase difference obtained from the time evolution of the real part of the phase.
\end{itemize}

\begin{table}[t]
	\caption{Initial condition.}
	\label{table:parameter}
	\centering
	\begin{tabular}{lc|lc}
	\hline
	Parameter & Value & Parameter & Value \\
	\hline
	$d$ & 1 & $\omega$ & 0 \\
	$\mathrm{Re}[\delta_{1}^{\uparrow}(0)]$ & $0$ & $\mathrm{Im}[\delta_{1}^{\uparrow}(0)]$ & $0$ \\
	$\mathrm{Re}[\delta_{1}^{\downarrow}(0)]$ & $0$ & $\mathrm{Im}[\delta_{1}^{\downarrow}(0)]$ & $0$ \\
	$\mathrm{Re}[\delta_{2}^{\uparrow}(0)]$ & $0$ & $\mathrm{Im}[\delta_{2}^{\uparrow}(0)]$ & $0$ \\
	$\mathrm{Re}[\delta_{2}^{\downarrow}(0)]$ & $0$ & $\mathrm{Im}[\delta_{2}^{\downarrow}(0)]$ & $0$ \\
	$\mathrm{Re}[\delta_{3}^{\uparrow}(0)]$ & $0$ & $\mathrm{Im}[\delta_{3}^{\uparrow}(0)]$ & $0$ \\
	$\mathrm{Re}[\delta_{3}^{\downarrow}(0)]$ & $0$ & $\mathrm{Im}[\delta_{3}^{\downarrow}(0)]$ & $0$ \\
	\hline
	\end{tabular}
\end{table}

\subsection{Results}
\begin{figure*}
\begin{center}
\includegraphics[width=0.8\linewidth]{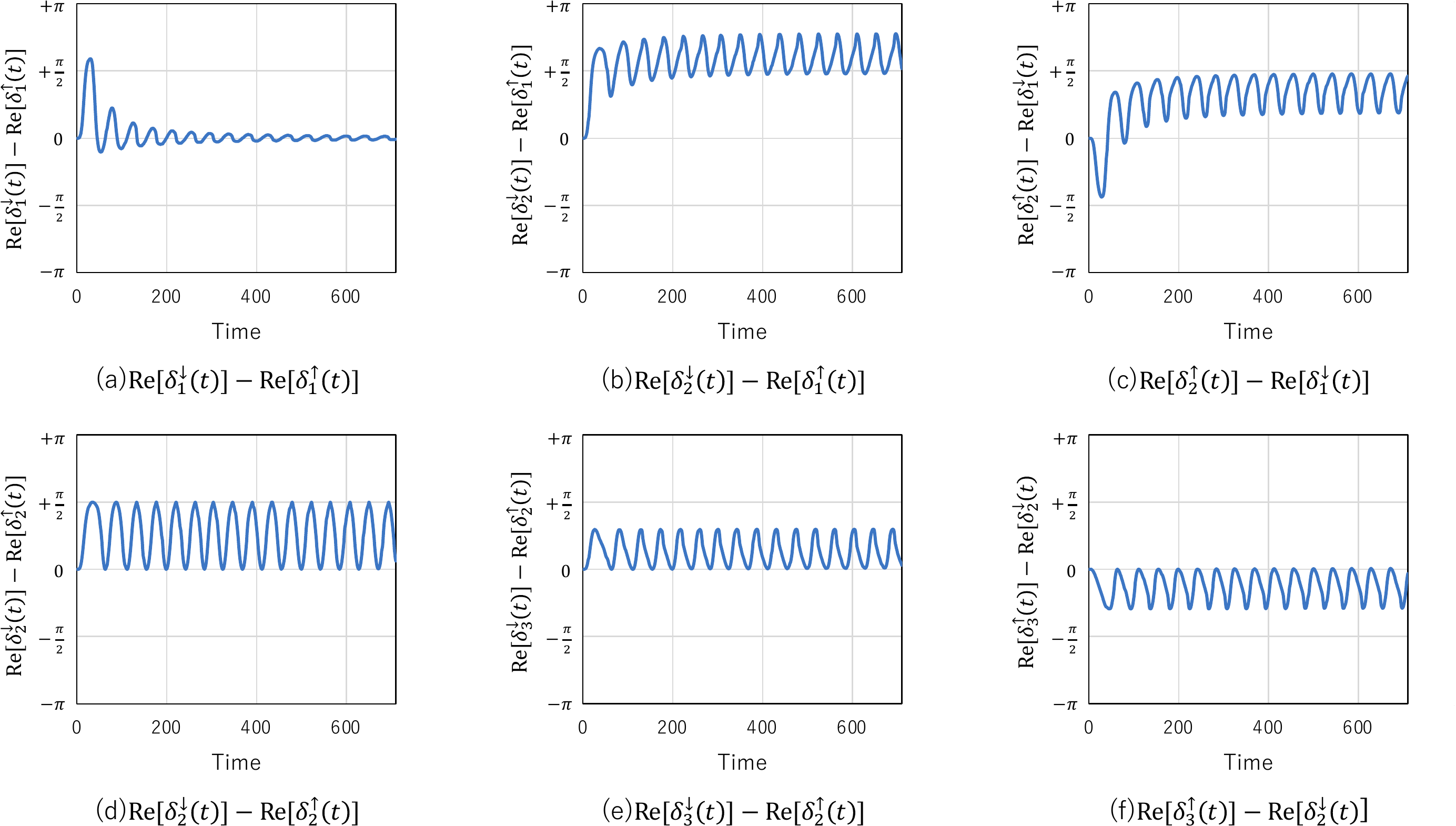}
\end{center}
\caption{Time evolution of the difference of real parts of phases. 
}
\label{fig:RealPhase}
\end{figure*}

Figure \ref{fig:RealPhase} shows the phase difference determined from the time evolution equation of the real part of the phase.
The horizontal axis shows the elapsed time, and the vertical axes in (a), (b), (c), (d), (e) and (f) plot 
$\mathrm{Re}[\delta_1^\downarrow] - \mathrm{Re}[\delta_1^\uparrow]$, $\mathrm{Re}[\delta_2^\downarrow] - \mathrm{Re}[\delta_1^\uparrow]$, $\mathrm{Re}[\delta_2^\uparrow] - \mathrm{Re}[\delta_1^\downarrow]$, $\mathrm{Re}[\delta_2^\downarrow] - \mathrm{Re}[\delta_2^\uparrow]$, $\mathrm{Re}[\delta_3^\downarrow] - \mathrm{Re}[\delta_2^\uparrow]$ and $\mathrm{Re}[\delta_3^\uparrow] - \mathrm{Re}[\delta_3^\downarrow]$, respectively.
Comparing the phase differences in Fig.~\ref{fig:RealPhase}, we can classify them into two types: phase difference 4(a), in which the oscillations are damped and converge to a specific value, and phase difference 4(b), 4(c), 4(d), 4(e), and 4(f), in which the oscillations are not damped and do not converge.

First, we analyze from the phase differences associated with the time evolution equation (\ref{eq:Red1u}), i.e., Figs.~4(a) and 4(b).
If the sine functions in the right-hand side of this equation cause Kuramoto-like synchronization, the first term on the right-hand side converges on
\[
\mathrm{Re}[\delta_{1}^{\downarrow}] - \mathrm{Re}[\delta_{1}^{\uparrow}] = +\frac{\pi}{2}.
\]
Similarly, the second term on the right-hand side converges on
\[
\mathrm{Re}[\delta_{2}^{\downarrow}] - \mathrm{Re}[\delta_{1}^{\uparrow}] = +\pi.
\]
Figure 4(a) shows the phase difference converging on $0$, which is different from the value ``$+\pi/2$'' predicted by the Kuramoto model.
In addition, Figure 4(b) shows that the phase difference does not converge.

Next, we examine the phase differences associated with the time evolution equation (\ref{eq:Red1d}), i.e., Figs.~4(a) and 4(c).
If the sine functions on the right-hand side of this equation cause Kuramoto-like synchronization, the first term on the right-hand side converges on 
\[
\mathrm{Re}[\delta_1^\uparrow] - \mathrm{Re}[\delta_1^\downarrow] =
-(\mathrm{Re}[\delta_1^\downarrow] - \mathrm{Re}[\delta_1^\uparrow]) = +\frac{\pi}{2}.
\]
Similarly, the second term on the right-hand side converges on
\[
\mathrm{Re}[\delta_2^\uparrow] - \mathrm{Re}[\delta_1^\downarrow] = 0.
\]
Figure 4(a) shows that the phase difference converges on $0$, which is different from the value ``$-\pi/2$'' predicted by the Kuramoto model.
In addition, Figure 4(c) shows that the phase difference does not converge.
The reason why the synchronization by the Kuramoto model does not occur is that the sine functions of the first terms on the right-hand side of (\ref{eq:Red1u}) and (\ref{eq:Red1d}) disturb phase synchronization. 

Note that, if Kuramoto-like synchronization does occur, equations (\ref{eq:Imd1u}) and (\ref{eq:Imd1d}) show that the amplitude increases exponentially, which alters the evolution of the imaginary part of the phase.
Although there is no Kuramoto-like synchronization here, it is necessary to investigate how the amplitude is affected through the behavior of the imaginary part of the phase. 

Since the coefficient of the sine functions on the right-hand sides of (\ref{eq:Imd1u}) and (\ref{eq:Imd1d}) is always positive, the sign of the sine functions determines whether the imaginary part of the phase increases or decreases with time.
Therefore, we examine the sign of the sine functions in (\ref{eq:Imd1u}) and (\ref{eq:Imd1d}) using the difference of the real part of the phase as shown in Fig.~4.

First, we examine the sign of the sine function on the right-hand side of (\ref{eq:Imd1u}).
The phase differences contributing to the sine function in the right-hand side of (\ref{eq:Imd1u}) are (a) and (b) in Fig.~4.
In Fig.~4(a), the phase difference in the real part converges on
$\mathrm{Re}[\delta_{1}^{\downarrow}] - \mathrm{Re}[\delta_{1}^{\uparrow}] \rightarrow 0$ with damped oscillations.
In this case, the sine function of the first term on the right-hand side of the corresponding equation (\ref{eq:Imd1u}) can be written as 
\[
\sin\left(\mathrm{Re}[\delta_{1}^{\downarrow}] - \mathrm{Re}[\delta_{1}^{\uparrow}] + \pi\right) \rightarrow \sin \pi = 0.
\]
Therefore, the sine function of the first term on the right-hand side of (\ref{eq:Imd1u}) does not contribute to the increase or decrease of $\mathrm{Im}[\delta^\uparrow_1(t)]$.
On the other hand, in Fig. 4(b), the phase difference of the real part does not converge but oscillates within a certain range over time.
Note that 
the minimum value is 
\[
\min\left(\mathrm{Re}[\delta_{2}^{\downarrow}(t)]-\mathrm{Re}[\delta_{1}^{\uparrow}(t)]\right) < \frac{\pi}{2},
\]
but is very close to $\pi/2$.
Details about how ``close'' will be given later.
The maximum value is 
\[
\max\left(\mathrm{Re}[\delta_{2}^{\downarrow}(t)]-\mathrm{Re}[\delta_{1}^{\uparrow}(t)]\right) \simeq \frac{4\pi}{5}. 
\]
Then, the sine function of the second term on the right-hand side of the corresponding equation (\ref{eq:Imd1u}) can be written as 
\begin{align*}
 \sin\left(\min\left(\mathrm{Re}[\delta_{2}^{\downarrow}(t)]-\mathrm{Re}[\delta_{1}^{\uparrow}(t)]\right)+\frac{\pi}{2}\right) &> 0,\\
 \sin\left(\max\left(\mathrm{Re}[\delta_{2}^{\downarrow}(t)]-\mathrm{Re}[\delta_{1}^{\uparrow}(t)]\right)+\frac{\pi}{2}\right) &< 0.
\end{align*}
Therefore, the sign of the sine function changes with time.
As a result, the temporal change in $\mathrm{Im}[\delta^\uparrow_1(t)]$ shows a repeated increase and decrease behavior cycle.

Next, we investigate the sign of the sine functions on the right-hand side of (\ref{eq:Imd1d}).
The phase differences contributing to the sine functions on the right-hand side of (\ref{eq:Imd1d}) are (a) and (c) in Fig.~4.
Figure 4(a) shows a damped oscillation that converges to $\mathrm{Re}[\delta_{1}^{\uparrow}] - \mathrm{Re}[\delta_{1}^{\downarrow}] \rightarrow 0$. 
In this case, the first term of the right-hand side the corresponding equation, (\ref{eq:Imd1d}), can be written as
\[
\sin\left(\mathrm{Re}[\delta_1^\uparrow] - \mathrm{Re}[\delta_1^\downarrow] -\pi\right) \rightarrow \sin(-\pi)=0.
\]
Therefore, the sine function of the first term of the right-hand side of (\ref{eq:Imd1d}) does not contribute to the increase or decrease of $\mathrm{Im}[\delta^\downarrow_1(t)]$.
On the other hand, in Fig. 4(c), the phase difference of the real part does not converge, but oscillates within a certain range over time.
The range with minimum value close to $\pi/5$ is written as follows
\[
\min\left(\mathrm{Re}[\delta_{2}^{\uparrow}(t)]-\mathrm{Re}[\delta_{1}^{\downarrow}(t)]\right) \simeq \frac{\pi}{5}.
\]
The maximum value is written as follows
\[
\max\left(\mathrm{Re}[\delta_{2}^{\uparrow}(t)]-\mathrm{Re}[\delta_{1}^{\downarrow}(t)]\right) < \frac{\pi}{2}.
\]
Then, the range of values of the sine function in the second term of the right-hand side of the corresponding equation, (\ref{eq:Imd1d}), is always negative, that is
\[
\sin\left(\left(\mathrm{Re}[\delta_{2}^{\downarrow}(t)]-\mathrm{Re}[\delta_{1}^{\uparrow}(t)]\right)-\frac{\pi}{2}\right) < 0.
\]
Thus, although the right-hand side of (\ref{eq:Imd1d}) differs from the phase synchronization of the Kuramoto model, the second term on the right-hand side of (\ref{eq:Imd1d}) produces a tendency for $\mathrm{Im}[\delta_{1}^{\downarrow}(t)]$ to decrease, since the second term is always negative.

\begin{figure*}
\begin{center}
\includegraphics[width=0.8\linewidth]{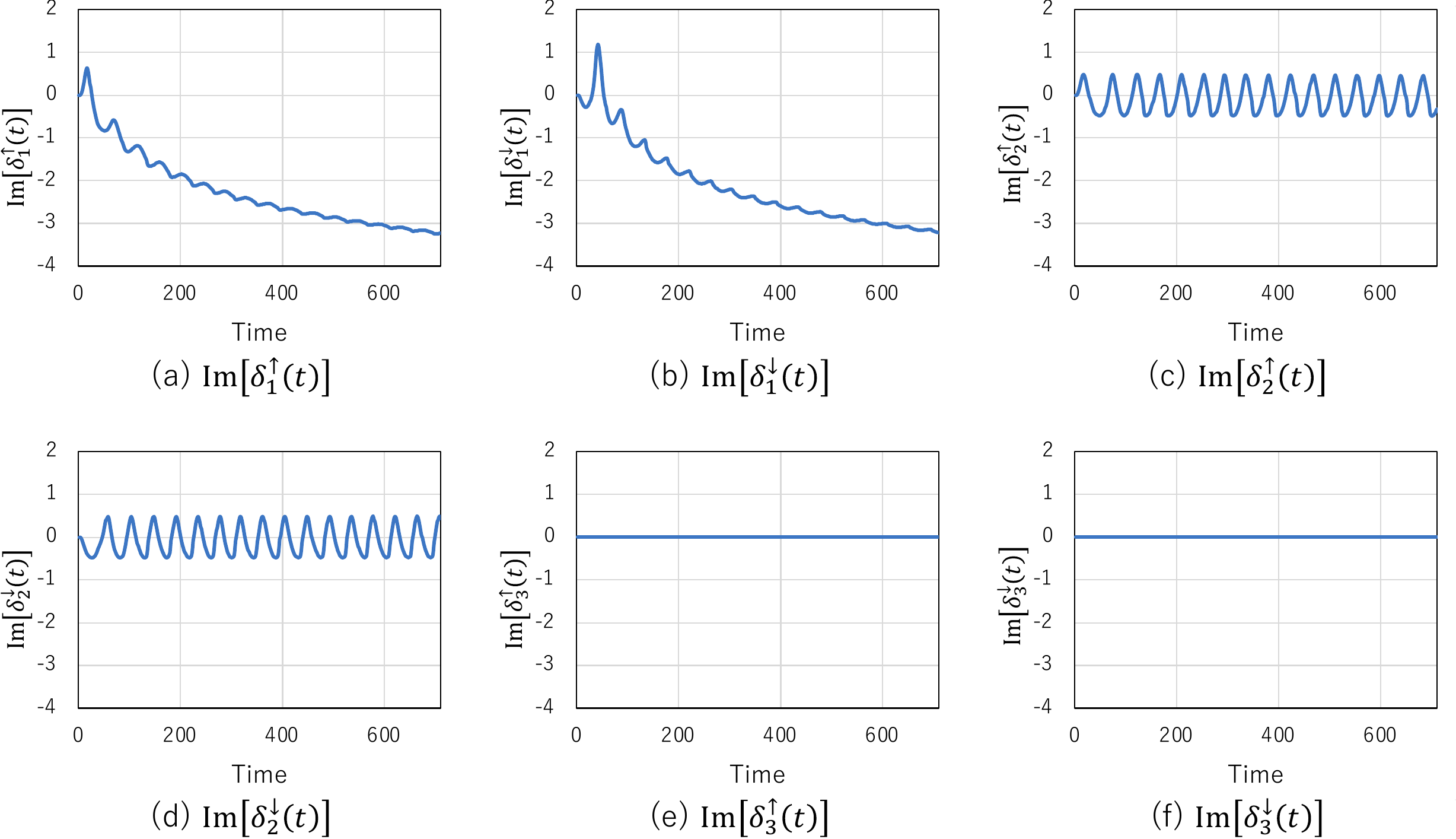}
\end{center}
\caption{Time evolution of the imaginary part of the phase}
\label{fig:3}
\end{figure*}

Figure 5 shows the time evolution for the imaginary part of the phases.
In Fig.~5, (a), (b), (c), (d), (e) and (f) represent  $\mathrm{Im}[\delta_{1}^{\uparrow}(t)]$,  $\mathrm{Im}[\delta_{1}^{\downarrow}(t)]$, $\mathrm{Im}[\delta_{2}^{\uparrow}(t)]$,  $\mathrm{Im}[\delta_{2}^{\downarrow}(t)]$,  $\mathrm{Im}[\delta_{3}^{\uparrow}(t)]$, and  $\mathrm{Im}[\delta_{1}^{\downarrow}(t)]$, respectively.
Comparing the phase differences in Fig.~\ref{fig:3}, we can classify them into two types; 5(a) and 5(b), in which the values tend to decrease, and 5(c), 5(d), 5(e), and 5(f), in which the values keep within certain ranges.
We want to investigate the existence of oscillation mode divergence, so we analyze $\mathrm{Im}[\delta_{1}^{\uparrow}(t)]$ and $\mathrm{Im}[\delta_{1}^{\downarrow}(t)]$ in Figs.~5(a) and (b).

The graphs in (a) $\mathrm{Im}[\delta_{1}^{\uparrow}(t)]$ and (b) $\mathrm{Im}[\delta_{1}^{\downarrow}(t)]$ have two common features.
The first is the presence of spikes that are periodic and whose height decays.
The second is the gradual global decrease in $\mathrm{Im}[\delta_{1}^{\uparrow}(t)]$ and $\mathrm{Im}[\delta_{1}^{\downarrow}(t)]$.
The spikes occur in opposite phases in the two graphs, and the global decreasing trend is almost identical in both graphs.

We analyze the occurrence of the spikes.
The occurrence of spikes indicates that the sign of the rate of change of $\mathrm{Im}[\delta_{1}^{\uparrow}(t)]$ and $\mathrm{Im}[\delta_{1}^{\downarrow}(t)]$ changes.
In the right-hand side of (\ref{eq:Imd1u}) and (\ref{eq:Imd1d}), there are two factors that cause the sign to change.
The first is the effect of the first term.
This effect can be seen from Fig.~4(a), which represents the oscillations that decay in the opposite phase.
The second is the effect of the second term. 
However, in the second term on the right-hand side, only (\ref{eq:Imd1u}) changes sign, (\ref{eq:Imd1d}) does not.
Considering the fact that the spikes appear in both Figs.~5(a) and 5(b) in common, and that they have the characteristic of attenuating in the opposite phase, we can recognize that this is the effect of the first term on the right-hand side.

\begin{figure*}
\begin{center}
\includegraphics[width=0.5\linewidth]{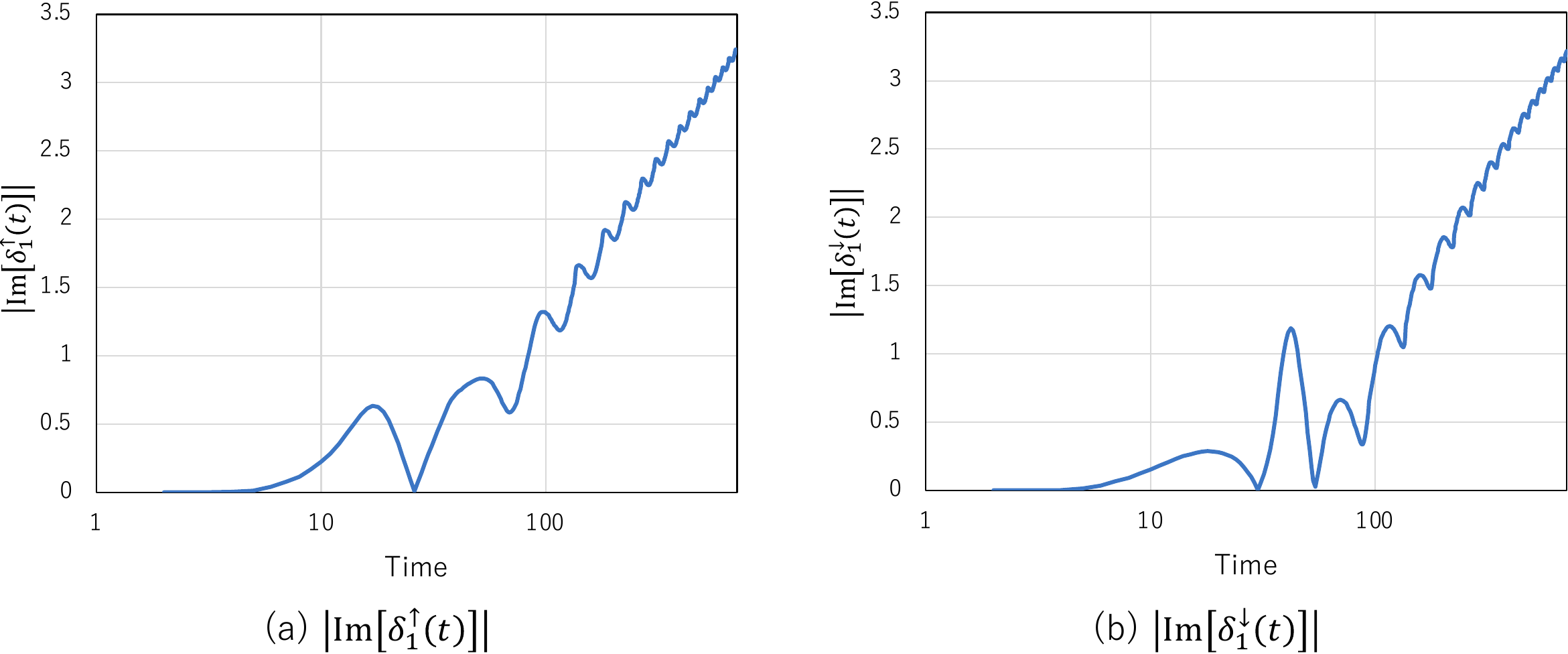}
\end{center}
\caption{Time evolution of the imaginary part of phase with respect to the time (logarithmic scale) }
\label{fig:4}
\end{figure*}
%

Next, we analyze the nature of the gradual global decrease in $\mathrm{Im}[\delta_{1}^{\uparrow}(t)]$ and $\mathrm{Im}[\delta_{1}^{\downarrow}(t)]$ to negative values.
The first terms on the right-hand side of (\ref{eq:Imd1u}) and (\ref{eq:Imd1d}) do not contribute to the global characteristics of the phenomena, because they represent the property of damping with short periodicity.
Therefore, this can be due to the effect of the second term on the right-hand side of (\ref{eq:Imd1u}) and (\ref{eq:Imd1d}).
While the sine function of the second right-hand side term in (\ref{eq:Imd1u}) has a sign change, the sine function of the second right-hand side term in (\ref{eq:Imd1d}) has no sign change and always takes a negative value.
As can be seen from Fig.~4(b), the sine function of the second term on the right-hand side in (\ref{eq:Imd1u}) is very close to $0$, even if a sign change occurs.
Thus, the fact that Figs.~5(a) and (b) globally behave in similar fashion means that the sine function of the second term on the right-hand side in (\ref{eq:Imd1d}) is not a factor that causes a significant difference.
Next, we focus on the coefficients before the sine function in (\ref{eq:Imd1u}) and (\ref{eq:Imd1d}).
The coefficients of the sine function in (\ref{eq:Imd1u}) and (\ref{eq:Imd1d}) are $\e^{-\mathrm{Im}[\delta_{2}^{\downarrow}(t)]+\mathrm{Im}[\delta_{1}^{\uparrow}(t)]}$ and $\e^{-\mathrm{Im}[\delta_{1}^{\downarrow}(t)]+\mathrm{Im}[\delta_{2}^{\uparrow}(t)]}$.
If $\mathrm{Im}[\delta_{1}^{\uparrow}(t)]$ and $\mathrm{Im}[\delta_{1}^{\downarrow}(t)]$ continue to decrease at negative values, the effect of the second term on the right-hand side weakens.
Therefore, we consider that the coefficients in Figs.~5(a) and (b) contribute to the gradual decrease in the global scale.

The observation drawn from Figs.~5(a) and (b) does not clarify whether the decrease of the imaginary part of the phase is bounded or not.
Therefore, we further investigate the characteristics of the time variation of $\mathrm{Im}[\delta_{1}^{\uparrow}(t)]$ and $\mathrm{Im}[\delta_{1}^{\downarrow}(t)]$ in more detail.
Figures~6(a) and (b) show the time evolution (logarithmic scale) of $|\mathrm{Im}[\delta_{1}^{\uparrow}(t)]|$ and $|\mathrm{Im}[\delta_{1}^{\downarrow}(t)]|$.
We confirm from Figs.~6(a) and (b) that $|\mathrm{Im}[\delta_{1}^{\uparrow}(t)]|$ and $|\mathrm{Im}[\delta_{1}^{\downarrow}(t)]|$ increase linearly, i.e., the absolute values of the imaginary parts of the two phases diverge logarithmically with time.
Therefore, we can conclude that $\mathrm{Im}[\delta_{1}^{\uparrow}(t)]$ and $\mathrm{Im}[\delta_{1}^{\downarrow}(t)]$ diverge, taking negative values.

By substituting the divergence of the imaginary components of the phase shown in Fig.~6(a) and (b) into (\ref{eq:PsiDeltaSol1}) and (\ref{eq:PsiDeltaSol2}), we can see that the state of the oscillation mode diverges.
From the above result, the logarithmic divergence of the imaginary part of the phases causes divergence in the strength of user dynamics to occur when divided communities experience correlation.

\section{Conclusions}
In this paper, to illustrate the mechanism of flaming phenomena in more detail, we investigated the characteristics of the user dynamics generated by the interaction of network oscillation modes, which are degenerate with eigenvalue $0$ and influence each other in Jordan canonical form.
We showed theoretically that the fundamental equations of the real network of degenerate modes with eigenvalue $0$ are equal to the unitary-transformed fundamental equation of networks, and so can cause flaming phenomena.
Numerical simulations demonstrated that divergence could occur in a model representing the coupling of oscillation mode $0$ instances.

The situation in which the eigenvalue $0$ instances couple as a Jordan block corresponds to the situation in which a fragmented community is coupled and the fragmentation of the network is resolved.
In general, changes in the network structure also have a significant impact on the graph spectrum, but in this work we focus only on changes where the degeneration of eigenvalue $0$ is resolved, that is, we consider only limited changes in the network structure.
Thus, even if coupling of fragmented communities does occur, it is possible that we are considering only the case wherein the coupling is relatively weak.
Degenerate eigenvalues are extremely uncommon and unlikely to occur in OSNs, with the exception of $0$ eigenvalue degeneracy. 
This degeneracy actually occurs because of the phenomenon related to the fragmentation of the network structure.
As mentioned in the Introduction, weak ties between fragmented communities that do not usually closely interact can cause significant changes in user dynamics. 
This is an interesting subject for further research.
The mechanism of flaming phenomena discussed in this paper corresponds to an example of user dynamics when weak connections occur between divided communities, and it is an example of user dynamics related to ``weak ties.'' Therefore, our results can be useful in modeling interesting user dynamics created by weak ties.

\appendix 
Let us consider the square of the following matrix: 
\begin{align}
\scalebox{0.7}{$\displaystyle
\left[
	\begin{array}{cccccc}
	0 & 1 & 0 & 0 & 0 & 0 \\
	0 & 0 & 1 & 0 & 0 & 0 \\
	0 & 0 & 0 & 0 & 0 & 0 \\
	0 & 0 & 0 & \omega_1 & 0 & 0\\
	0 & 0 & 0 & 0 & \omega_2 & 0\\
	0 & 0 & 0 & 0 & 0 & \ddots
	\end{array}
  \right] $} ^{2}
  =
\scalebox{0.7}{$\displaystyle
\left[
	\begin{array}{cccccc}
	0 & 0 & 1 & 0 & 0 & 0 \\
	0 & 0 & 0 & 0 & 0 & 0 \\
	0 & 0 & 0 & 0 & 0 & 0 \\
	0 & 0 & 0 & \omega_1^2 & 0 & 0\\
	0 & 0 & 0 & 0 & \omega_2^2 & 0\\
	0 & 0 & 0 & 0 & 0 & \ddots
	\end{array}
  \right] $} . \notag
\end{align}
By arranging rows and columns, we have the following matrix, which is the Jordan canonical form:
\begin{align}
\scalebox{0.7}{$\displaystyle
\left[
	\begin{array}{cccccc}
	0 & 1 & 0 & 0 & 0 & 0 \\
	0 & 0 & 0 & 0 & 0 & 0 \\
	0 & 0 & 0 & 0 & 0 & 0 \\
	0 & 0 & 0 & \omega_1^2 & 0 & 0\\
	0 & 0 & 0 & 0 & \omega_2^2 & 0\\
	0 & 0 & 0 & 0 & 0 & \ddots
	\end{array}
  \right] $} . \notag
\end{align}

\section*{Acknowledgment}
This research was supported by Grant-in-Aid for Scientific Research (B) No.~19H04096 (2019--2021) and No.~20H04179 (2020--2022), and Grant-in-Aid for Scientific Research (C) No.~18K11271 (2018--2020)  from the Japan Society for the Promotion of Science (JSPS).


\end{document}